\documentclass[12pt]{article}
\usepackage{epsfig}
\begin{document}
{\large \sf

~~\vspace{1cm}
\begin{center}

{\LARGE \sf
A New Approach to Solve the Low-lying States of\\

\vspace{.2cm}

the Schroedinger Equation\footnote{{\normalsize \sf This research
was supported in
part by the U.S. Department of Energy Grant DE-FG02-92ER-40699}}}\\

\vspace{2cm}

{\Large \sf T. D. Lee}\\

\vspace{1cm}

{\large \it Physics Department, Columbia University, New York, NY 10027}\\

\vspace{.2cm}

{\large \it China Center of Advanced Science and Technology },\\

\vspace{.2cm}

{\large \it (CCAST-World Laboratory), P.O. Box 8730,}\\

\vspace{.2cm}

{\large \it Beijing 100080, People's Republic of China}

\vspace{1cm}

{\Large \bf \sf Abstract}

\end{center}

{\large \sf

We review a new iterative procedure to solve the low-lying states
of the Schroedinger equation, done in collaboration with Richard
Friedberg. For the groundstate energy, the $n^{th}$ order
iterative energy is bounded by a finite limit, independent of $n$;
thereby it avoids some of the inherent difficulties faced by the
usual perturbative series expansions. For a fairly large class of
problems, this new procedure can be proved to give convergent
iterative solutions. These convergent solutions include the long
standing difficult problem of a quartic potential with either
symmetric or asymmetric minima.

}

\newpage

\section*{\large \bf 1. Importance of Low-lying States}
\setcounter{section}{1}
\setcounter{equation}{0}

It is a common belief that most of natural phenomena should be
described by solutions of the Schroedinger equation. Take the
nonrelativistic Schroedinger equation of a system of charged
particles with Coulomb interactions. In principle, this single
N-dimensional second order linear partial differential equation
contains in its solutions all information about many disciplines:
atomic and molecular physics (except for small relativistic
corrections), condensed matter physics, chemistry and biology.
Furthermore, for most of these applications we need only knowledge
about its low-lying states. For these low-lying states, actually
we do have very good information about their wave functions when
the Coulomb interaction is strong; i.e., when the wave amplitude
is large. How come that we still have great difficulties in
solving these equations? Quite often, the challenge lies in how to
handle the large number of configurations when the interaction is
weak.

In quantum chromodynamics (QCD) and quantum electrodynamics (QED)
the tasks are even more difficult since we do not have any exact
solution. Besides the very promising lattice and other numerical
calculations, we rely mostly on perturbative expansions. Yet, such
expansion often leads to a divergent series with zero radius of
convergence, as exemplified by the perturbative calculation of the
gyromagnetic ratio g of the muon[1-3]:
\begin{eqnarray}\label{e1.1}
\frac{1}{2}(g-2)_\mu &=& \frac{1}{2}\frac{\alpha}{\pi} +
0.765857376(\frac{\alpha}{\pi})^2 \nonumber\\
&& + 24.05050898(\frac{\alpha}{\pi})^3 +
131.0(\frac{\alpha}{\pi})^4 \nonumber\\
&& + 677(40)(\frac{\alpha}{\pi})^5 + \cdots
\end{eqnarray}
where $\alpha$ is the fine-structure constant. The coefficients
are all positive, and each successive one becomes larger and
larger. There are good reasons [4] to believe that the vacuum
state of QED would be unstable against pair creations if $\alpha$
could be analytically continued to the negative region. Thus, the
difficulty of perturbative series seems to be again closely
related to our inability to solve the low-lying states of the
corresponding Schroedinger equation. In QCD, because of the
existence of instanton configurations, similar serious problems
also exist for its vacuum and other low-lying states.

This situation may be illustrated by the following one dimensional
problem with a quartic potential of degenerate minima. The
Schroedinger equation can be written as
\begin{eqnarray}\label{e1.2}
(T+V-E)\psi = 0
\end{eqnarray}
where $T=-\frac{1}{2}\frac{d^2}{dx^2}$ and
\begin{eqnarray}\label{e1.3}
V(x)=\frac{1}{2} g^2(x^2-a^2)^2.
\end{eqnarray}
An alternative form of the same problem may be obtained by setting
$q\equiv \sqrt{2ga}(a-x)$, so that the Hamiltonian becomes
\begin{eqnarray}\label{e1.4}
H=T+V \equiv 2ga {\cal H}
\end{eqnarray}
with
\begin{eqnarray}\label{e1.5}
{\cal H}=-\frac{1}{2}\frac{d^2}{dq^2}+\frac{1}{2} q^2(1-e~q)^2,
\end{eqnarray}
in which ${\cal H}$, $q$ and $e=1/ \sqrt{8ga^3}$ are all
dimensionless. The perturbative expression of the groundstate
energy $E$ is[5-14]
\begin{eqnarray}\label{e1.6}
E=ga-\frac{1}{4a^2}-\frac{9}{64}\frac{1}{ga^5}-\frac{89}{512}\frac{1}{g^2a^8}
-O(\frac{1}{g^3a^{11}}),
\end{eqnarray}
or, in terms of the dimensionless coupling $e^2$
\begin{eqnarray}\label{e1.7}
E=2ga[\frac{1}{2}-e^2-\frac{9}{2}e^4-\frac{89}{2}e^6-\cdots],
\end{eqnarray}
which has a similar characteristics as the $(g-2)$ expansion
(\ref{e1.1}).

In this one-dimensional case, from (\ref{e1.5}), one sees that for
$e^2$ negative, ${\cal H}$ has no groundstate; this explains why
the power series (\ref{e1.7}) is divergent. For $e^2=1/8ga^3$
positive, it is also possible to trace the origin of the
difficulty. The corresponding perturbative series of the
groundstate wave function $\psi(x)$ may be written as
\begin{eqnarray}\label{e1.8}
\psi(x)=e^{-gS_0-S_1-g^{-1}S_2-g^{-2}S_3-\cdots}~.
\end{eqnarray}
Since the potential $V(x)$ has two degenerate minima at $x=\pm a$,
there are two different expansions depending on the normalization
condition:
\begin{eqnarray}\label{e1.9}
\psi(a)=1,~~~~{\sf or}~~~~\psi(-a)=1.
\end{eqnarray}
In either case, the perturbative series is divergent for any $g$.
However, for $g$ large, the wave function $\psi(x)$ can be
approximately described by the first two terms of these
expansions. For $x>0$, using the normalization $\psi(a)=1$, we
find
\begin{eqnarray}\label{e1.10}
e^{-gS_0-S_1}=\frac{2a}{a+x}e^{-\frac{g}{3}(a-x)^2(2a+x)},
\end{eqnarray}
and for $x<0$ with $\psi(-a)=1$,
\begin{eqnarray}\label{e1.11}
e^{-gS_0-S_1}=\frac{2a}{a-x}e^{-\frac{g}{3}(a+x)^2(2a-x)}.
\end{eqnarray}
Unlike $\psi(x)$, neither expression has a zero derivative at
$x=0$. Furthermore, each carries a spurious pole-term, at $x=-a$
for (\ref{e1.10}) and $x=a$ for (\ref{e1.11}). In addition, if one
stays with the perturbative series (\ref{e1.8}), each successive
higher order expansion accentuates further these spurious pole
terms. This suggests that we must not follow the high order
perturbative expressions, especially in the region $x$ near $0$,
when $\psi(x)$ is exponentially small.

In all these problems, it is not difficult to construct a trial
function $\phi$ such that
\begin{eqnarray}\label{e1.12}
\phi \approx \psi~~~~~~~{\sf when}~~~\psi~~~{\sf is~~large},
\end{eqnarray}
and with the correct symmetry properties of $\psi$ (e.g.,
$\phi(x)=\phi(-x)$ for the quartic potential problem
(\ref{e1.2})-(\ref{e1.3})). The question is how to device a
convergent procedure that can lead from the trial function $\phi$
to the correct $\psi$. In the following we will discuss such a
method.

\newpage

\section*{\large \bf 2. The New Method}
\setcounter{section}{2} \setcounter{equation}{0}

As examples of the types of Schroedinger equations that we are
interested, consider the bosonic component of QED or QCD in the
axial gauge, or a system of nonrelativistic particles. In either
of these cases the kinetic energy $T$ is a quadratic function of
momentum operators. Through a linear transformation of the
relevant coordinate variables, the Hamiltonian $H$ can be written
as
\begin{eqnarray}\label{e2.1}
H = T + V(q)
\end{eqnarray}
where
\begin{eqnarray}\label{e2.2}
q =(q_1,~q_2,~\cdots,~q_N)
\end{eqnarray}
and
\begin{eqnarray}\label{e2.3}
T = -\frac{1}{2} \sum\limits_{i=1}^{N} \partial^2/\partial q_i^2 =
-\frac{1}{2} \nabla^2,
\end{eqnarray}
with $N=\infty$ in the case of a field theory. Let $\psi(q)$ be
the groundstate determined by
\begin{eqnarray}\label{e2.4}
(H-E)\psi = 0.
\end{eqnarray}
To derive $\psi(q)$, we proceed as follows [15-19]:

\noindent 1. Construct a good trial function $\phi(q)$, so that
\begin{eqnarray}\label{e2.5}
\phi(q)\approx \psi(q)
\end{eqnarray}
in regions when $\psi(q)$ is expected to be large. An example of
how to construct such a trial function is given in Appendix A.

\noindent 2. Construct $U(q)-E_0$ by differentiating $\phi(q)$:
\begin{eqnarray}\label{e2.6}
U(q)-E_0 \equiv (\frac{1}{2} \nabla^2 \phi)/\phi.
\end{eqnarray}
Define
\begin{eqnarray}\label{e2.7}
H_0 = -\frac{1}{2} \nabla^2 + U(q).
\end{eqnarray}
We have
\begin{eqnarray}\label{e2.8}
(H_0-E_0)\phi(q) = 0
\end{eqnarray}
where the constant $E_0$ may be defined by requiring, say, the
potential energy part $U(q)$ of $H_0$ to satisfy at $q =\infty$
\begin{eqnarray}\label{e2.9}
U(\infty)= 0.
\end{eqnarray}
Alternatively, in problems like the quartic potential
(\ref{e1.3}), the constant $E_0$ can be more conveniently
determined by requiring
\begin{eqnarray}\label{e2.10}
{\sf minimum~~of}~~U(q)=0.
\end{eqnarray}
(Since our method deals only with the operator $H_0-E_0$, it is
independent of the particular way to define $E_0$.) Introduce the
differences $h$ and ${\cal E}$ by
\begin{eqnarray}\label{e2.11}
H_0-H = U(q) - V(q) \equiv h(q)
\end{eqnarray}
and
\begin{eqnarray}\label{e2.12}
E_0-E\equiv {\cal E}.
\end{eqnarray}
The original Schroedinger equation (\ref{e2.4}) can be written as
\begin{eqnarray}\label{e2.13}
(H_0-E_0)\psi(q)=(h- {\cal E})\psi(q).
\end{eqnarray}
For the discussions of the groundstate, we take both
\begin{eqnarray}\label{e2.14}
\psi(q)~~~{\sf and}~~~\phi(q)~~~{\sf to~be~real~and~positive}.
\end{eqnarray}

Multiplying (\ref{e2.13}) by $\phi(q)$ and (\ref{e2.8}) by
$\psi(q)$, then taking their difference, we have
\begin{eqnarray}\label{e2.15}
-\frac{1}{2} \nabla \cdot (\phi \nabla \psi - \psi \nabla \phi) =
(h-{\cal E})\phi~\psi.
\end{eqnarray}
The integration of (\ref{e2.15}) over all space gives $\int
(h-{\cal E})\phi \psi d^Nq = 0$; therefore,
\begin{eqnarray}\label{e2.16}
{\cal E}= \int h~\phi~\psi~d^N q~/~\int \phi~\psi~d^N q.
\end{eqnarray}

It is convenient to introduce
\begin{eqnarray}\label{e2.17}
f(q)= \psi(q)/\phi(q);
\end{eqnarray}
(\ref{e2.16}) becomes then
\begin{eqnarray}\label{e2.18}
{\cal E}= \int h(q) \phi^2(q) f(q) d^N q~\bigg/~\int \phi^2(q)
f(q) d^N q.
\end{eqnarray}

\noindent 3. We propose to solve the Schroedinger equation
(\ref{e2.13}) by an iterative process, defined by
\begin{eqnarray}\label{e2.19}
(H_0-E_0)\psi_n(q)=(h(q)- {\cal E}_n)\psi_{n-1}(q)
\end{eqnarray}
with $n=1,~2,~\cdots$ and for $n=0$,
\begin{eqnarray}\label{e2.20}
\psi_0(q)=\phi(q).
\end{eqnarray}
Multiplying (\ref{e2.19}) by $\phi(q)$ and (\ref{e2.8}) by
$\psi_n(q)$, then taking their difference, we derive
\begin{eqnarray}\label{e2.21}
-\frac{1}{2} \nabla \cdot (\phi \nabla \psi_n - \psi_n \nabla
\phi) = (h-{\cal E}_n)\phi~\psi_{n-1}.
\end{eqnarray}
Analogous to (\ref{e2.16}), we have
\begin{eqnarray}\label{e2.22}
{\cal E}_n= \int h~\phi~\psi_{n-1} d^N q~\bigg/~\int
\phi~\psi_{n-1} d^N q.
\end{eqnarray}
Note that in each step of the iteration from $n-1$ to $n$, for any
particular solution $\psi^0_n(q)$ that satisfies (\ref{e2.19}), so
is
\begin{eqnarray}\label{e2.23}
\psi_n(q)=\psi^0_n(q) + c_n \phi(q)
\end{eqnarray}
also a solution where $c_n$ is an arbitrary constant. Since
$\phi(q)$ is positive in accordance with (\ref{e2.14}), we can
choose $c_n$ so that
\begin{eqnarray}\label{e2.24}
\psi_n(q)~~~{\sf is~positive~everywhere}.
\end{eqnarray}
Define
\begin{eqnarray}\label{e2.25}
f_n(q)= \psi_n(q)/\phi(q)
\end{eqnarray}
and therefore (\ref{e2.22}) becomes
\begin{eqnarray}\label{e2.26}
{\cal E}_n= \int h~\phi^2 f_{n-1} d^N q~\bigg/~\int \phi^2 f_{n-1}
d^N q.
\end{eqnarray}
From (\ref{e2.24}), we have
\begin{eqnarray}\label{e2.27}
f_n(q)~~~{\sf positive}.
\end{eqnarray}
As we shall see, a particularly convenient choice of the constant
$c_n$ is to set the minimum of $f_n(q)$ to be $1$ and therefore
\begin{eqnarray}\label{e2.28}
f_n(q)\geq 1.
\end{eqnarray}
From (\ref{e2.26}) and (\ref{e2.27}), it follows that ${\cal E}_n$
is bounded if $h(q)$ is bounded. Therefore, as $n \rightarrow
\infty $, $\lim{\cal E}_n$ cannot be $\infty$. This avoids the
type of divergence difficulties that appears in the example
(\ref{e1.7}). It is therefore reasonable to expect that
$\lim_{n\rightarrow\infty} {\cal E}_n$ should converge to the
correct ${\cal E}$, at least when $h$ is small. As we shall see,
for a fairly large class of problems, including the quartic
potential (\ref{e1.3}), the convergence of ${\cal E}_n$ turns out
to be independent of the magnitude of $h(q)$, provided that it is
finite and satisfies some general conditions.

In this paper, we concentrate on the groundstate. Extensions to
some low-lying states can be found in Ref.[17].

\newpage

\section*{\large \bf 3. An Electrostatic Analog}
\setcounter{section}{3} \setcounter{equation}{0}

In terms of the ratio $f_n=\psi_n/\phi$ introduced in
(\ref{e2.25}), the n$^{{\sf th}}$ order iterative equation
(\ref{e2.21}) can be written as
\begin{eqnarray}\label{e3.1}
-\frac{1}{2} \nabla \cdot (\phi^2 \nabla f_n) = \sigma_n(q)
\end{eqnarray}
where
\begin{eqnarray}\label{e3.2}
\sigma_n(q) \equiv (h(q)-{\cal E}_n)\phi^2(q) f_{n-1}(q)
\end{eqnarray}
and when $n=0$,
\begin{eqnarray}\label{e3.3}
f_0(q)=1
\end{eqnarray}
in accordance with (\ref{e2.20}). Assuming that $\psi_{n-1}(q)$
has already been solved, we can determine ${\cal E}_n$ through
(\ref{e2.22}). Therefore, $\sigma_n(q)$ is a known function.

Consider a dielectric medium with a q-dependent dielectric
constant, given by
\begin{eqnarray}\label{e3.4}
\kappa(q)\equiv \phi^2(q).
\end{eqnarray}
Interpret $\sigma_n(q)$ as an external electrostatic charge
distribution, $\frac{1}{2} f_n$ the electrostatic potential,
$-\frac{1}{2}\nabla f_n$ the electrostatic field and
\begin{eqnarray}\label{e3.5}
D_n \equiv -\frac{1}{2}\kappa \nabla f_n
\end{eqnarray}
the corresponding displacement vector field. Thus (\ref{e3.1})
becomes
\begin{eqnarray}\label{e3.6}
\nabla\cdot D_n =\sigma_n,
\end{eqnarray}
the Maxwell equation for this electrostatic analog problem.

At infinity, $\phi(\infty)=0$. In accordance with (\ref{e3.4}) -
(\ref{e3.5}), we also have $D_n(\infty)=0$. Hence the integration
of (\ref{e3.6}) leads to the total external electrostatic charge
to be also zero; i.e.,
\begin{eqnarray}\label{e3.7}
\int\sigma_n(q)~ d^Nq=0
\end{eqnarray}
which is the same result given by (\ref{e2.26}) for the
determination of ${\cal E}_n$. Because the dielectric constant
$\kappa(q)$ in this analog problem is zero at $q=\infty$, the
dielectric media becomes a perfect dia-electric at $\infty$. Thus,
the equation of zero total charge, given by (\ref{e3.7}), may
serve as a much simplified model of charge confinement, analogous
to color confinement in quantum chromodynamics.

We note that (\ref{e3.1}) can also be derived from a minimal
principle by defining
\begin{eqnarray}\label{e3.8}
I(f_n(q)) \equiv \int\{\frac{1}{4}\kappa(\nabla f_n)^2 - \sigma_n
f_n\}d^N q.
\end{eqnarray}
Because of (\ref{e3.7}), the functional $I(f_n(q))$ is invariant
under
\begin{eqnarray}\label{e3.9}
f_n(q) \rightarrow f_n(q) + {\sf constant}.
\end{eqnarray}
Since the quadratic part of $I(f_n(q))$ is the integral of the
positive definite $\frac{1}{4}\kappa(\nabla f_n)^2$, the curvature
of $I(f_n(q))$ in the functional space $f_n(q)$ is always
positive. Hence, $I(f_n(q))$ has a minimum, and that minimum
determines a unique electrostatic field $-\frac{1}{2}\nabla f_n$,
as we shall see. To establish the uniqueness, let us assume two
different $\nabla f_n$, both satisfy (\ref{e3.1}), with the same
$\kappa=\phi^2$ and the same $\sigma_n$; their difference would
then satisfy (\ref{e3.1}) with a zero external charge
distribution. For $\sigma_n=0$, the minimum of $I(f_n(q))$ is
clearly zero with the corresponding $\nabla f_n=0$. To derive
$f_n(q)$ from $\nabla f_n$, there remains an additive constant at
each iteration, as already noted in (\ref{e2.23}). As we shall
discuss in the next section, this arbitrariness allows us the
freedom to derive different types of convergent series.

We note that in this electrostatic analog, the problem depends
only on two input-functions: $\phi^2(q)$ and the product
\begin{eqnarray}\label{e3.10}
h(q)\phi^2(q).
\end{eqnarray}
The original potential $V(q)$ no longer appears explicitly.
Consider, e.g., a system of particles with electric charges. For
configurations when the Coulomb energy $V(q)$ is attractive and
near its singularity, we have good knowledge about the wave
amplitude $\psi(q)$, which is large but regular. The same should
be true for a good trial function $\phi(q)$. In accordance with
(\ref{e1.12}), we expect
\begin{eqnarray}\label{e3.11}
h(q)~~~{\sf small~~~~when}~\psi~{\sf is~large}.
\end{eqnarray}
For configurations when $\psi$ is small, so should be $\phi$.
Thus, a good trial function $\phi$ would result in a small
$h\phi^2$ everywhere even though $h$ may not.

When $N=1$, we may denote the coordinate (\ref{e2.2}) by a single
$x$. Correspondingly (\ref{e3.5}) and (\ref{e3.6}) become
\begin{eqnarray}\label{e3.12}
D_n=-\frac{1}{2}\kappa \frac{df_n}{dx}
\end{eqnarray}
and
\begin{eqnarray}\label{e3.13}
\frac{dD_n}{dx}=\sigma_n(x).
\end{eqnarray}
Since $D_n(\infty)=0$, we have
\begin{eqnarray}\label{e3.14}
D_n(x)=-\int\limits^\infty_x \sigma_n(z)dz
\end{eqnarray}
and therefore
\begin{eqnarray}\label{e3.15}
f_n(x)=f_n(\infty) - 2 \int\limits_{x}^{\infty}\phi^{-2}(y)dy
\int\limits_{y}^{\infty}  \sigma_n(z)dz
\end{eqnarray}
with
\begin{eqnarray}\label{e3.16}
\sigma_n(x)=(h(x)- {\cal E}_n) \phi^2(x)f_{n-1}(x),
\end{eqnarray}
which satisfies
\begin{eqnarray}\label{e3.17}
\int\limits^\infty_{-\infty} \sigma_n(x)dx = 0.
\end{eqnarray}
Therefore,
\begin{eqnarray}\label{e3.18}
D_n(-\infty)=D_n(\infty)=0
\end{eqnarray}
and
\begin{eqnarray}\label{e3.19}
f_n(\infty)-f_n(-\infty)=  2
\int\limits_{-\infty}^{\infty}\phi^{-2}(y)dy
\int\limits_{y}^{\infty} \sigma_n(z)dz.
\end{eqnarray}

When $N>1$, for radially symmetric problems we have
\begin{eqnarray}\label{e3.20}
h=h(r),~~~~~\phi^2=\kappa(r)
\end{eqnarray}
where
\begin{eqnarray}\label{e3.21}
r^2=q_1^2+q_2^2+\cdots+q_N^2.
\end{eqnarray}
Likewise $\psi$, $\psi_n$ and $\sigma_n$ are all functions of $r$
only. The radial component of $D_n$ is
\begin{eqnarray}\label{e3.22}
(D_n)_r=-\frac{1}{2}\kappa(r)\frac{df_n(r)}{dr}.
\end{eqnarray}
Correspondingly, (\ref{e3.6}) becomes
\begin{eqnarray}\label{e3.23}
\frac{1}{r^{N-1}}\frac{d}{d
r}\bigg(r^{N-1}(D_n)_r\bigg)=\sigma_n(r),
\end{eqnarray}
with
\begin{eqnarray}
\sigma_n(r)=(h(r)-{\cal E}_n)\phi^2(r)f_{n-1}(r).\nonumber
\end{eqnarray}
Hence,
\begin{eqnarray}\label{e3.24}
(D_n)_r=-\frac{1}{r^{N-1}}\int\limits^\infty_r
z^{N-1}\sigma_n(z)~dz
\end{eqnarray}
and on account of (\ref{e3.22}),
\begin{eqnarray}\label{e3.25}
f_n(r)=f_n(\infty)- \int\limits_{r}^{\infty}\frac{2~
dy}{y^{N-1}\phi^2(y)} \int\limits_{y}^{\infty}
z^{N-1}\sigma_n(z)~dz.
\end{eqnarray}
Assuming as $r\rightarrow \infty$,
\begin{eqnarray}\label{e3.26}
\phi^2(r) \rightarrow a~e^{-br^l}
\end{eqnarray}
with $a,~b$ and $l$ all positive constants. We assume
\begin{eqnarray}\label{e3.27}
l>2.
\end{eqnarray}
In accordance with (\ref{e3.16}), as $r\rightarrow \infty$
\begin{eqnarray}\label{e3.28}
\sigma_n(r) \rightarrow c~e^{-br^l}
\end{eqnarray}
where
\begin{eqnarray}\label{e3.29}
c=(h(\infty)-{\cal E}_n)~a~f_{n-1}(\infty).
\end{eqnarray}
By partial integrations, we see that as $r\rightarrow \infty$,
\begin{eqnarray}\label{e3.30}
f_n(r)\rightarrow f_n(\infty)- \frac{2}{b}~~\frac{(h(\infty)-{\cal
E}_n)f_{n-1}(\infty)}{l(l-2)
r^{l-2}}~\bigg[1+O(\frac{1}{~br^l})\bigg] .
\end{eqnarray}
When the dimension $N=1$, the same expression applies, with $r$
replaced by $x$.

Because of (\ref{e3.7}). in the $N>1$ radially symmetric case we
have
\begin{eqnarray}\label{e3.31}
\int\limits_{0}^{\infty}r^{N-1}\sigma_n(r)~dr=0;
\end{eqnarray}
therefore (\ref{e3.25}) is identical to
\begin{eqnarray}\label{e3.32}
f_n(r)=f_n(0)- \int\limits_0^{r}\frac{2~ dy}{y^{N-1}\phi^2(y)}
\int\limits_0^{y} z^{N-1}\sigma_n(z)~dz.
\end{eqnarray}
When $N=1$ and $V(x)$ is an even function of $x$, (\ref{e3.32})
applies with the replacement of $r$ by $x$; if $V(x)$ is not an
even function then we have, from (\ref{e3.15}) and  (\ref{e3.19}),
\begin{eqnarray}\label{e3.33}
f_n(x)=f_n(-\infty)- 2\int\limits_{-\infty}^{x} \phi^{-2}(y)dy
\int\limits_{-\infty}^{y}\sigma_n(z)~dz.
\end{eqnarray}

For an arbitrary $N>1$ dimensional radially non-symmetric problem,
the general solution of the electrostatic analog problem can no
longer be reduced to simple quadratures. Thus, numerical analysis
may become important. In that case, the minimal principle using
the functional $I(f(q))$ given by (\ref{e3.8}) may be of some
practical use.

\newpage

\section*{\large \bf 4. Hierarchy Theorem}
\setcounter{section}{4} \setcounter{equation}{0}

In this section we restrict our discussions to either (i) the
$N$-dimensional radially symmetric case in which the functions $V$
in (\ref{e2.1}) and $h$ in (\ref{e2.11}) are
\begin{eqnarray}\label{e4.1}
V=V(r)~~~{\sf and}~~~h=h(r)
\end{eqnarray}
where $r^2=q_1^2+q_2^2+\cdots+q_N^2$ as in (\ref{e3.21}), or (ii)
the one-dimensional case (\ref{e3.12})-(\ref{e3.17}) with even
functions
\begin{eqnarray}\label{e4.2}
V(x)=V(-x)~~~{\sf and}~~~h(x)=h(-x).
\end{eqnarray}
In the latter case, we need only to consider the region
\begin{eqnarray}\label{e4.3}
x\equiv r \geq 0.
\end{eqnarray}
Thus, we need only to consider (i), since the $N$-dimensional
radially symmetric solution (\ref{e3.25}) reduces to the
one-dimensional case (\ref{e3.15}), with $N=1$ and $r$ replaced by
$x$. Furthermore, we assume $h(r)$ to satisfy at all finite $r>0$,
\begin{eqnarray}\label{e4.4}
h(r)> 0~~{\sf and}~~h'(r)< 0;
\end{eqnarray}
at infinity,
\begin{eqnarray}\label{e4.5}
h(\infty)= 0.
\end{eqnarray}
Throughout the paper, ' denotes $\frac{d}{dr}$. An example of such
$h(r)$ is given in Appendix A for the quartic potential problem
(\ref{e1.3}).

In accordance with (\ref{e3.25}), each $n^{th}$ iterative solution
$f_n(r)$ carries an independent additive constant. In the
following, we differentiate two sets of sequences, labelled $A$
and $B$, satisfying different boundary conditions:
\begin{eqnarray}\label{e4.6}
f_n(\infty)=1~~{\sf for~~all}~~n,~~~{\sf in~~Case}~~(A)
\end{eqnarray}
or
\begin{eqnarray}\label{e4.7}
f_n(0)=1~~~~~{\sf for~~all}~~~n,~~~{\sf in~~Case}~(B).
\end{eqnarray}
Thus, by using (\ref{e3.25}) or (\ref{e3.32}) we have in Case
$(A)$
\begin{eqnarray}\label{e4.8A}
f_n(r)=1 - 2 \int_{r}^{\infty}y^{-N+1}\phi^{-2}(y)dy
\int_{y}^{\infty}z^{N-1}\phi^{2}(z) (h(z)- {\cal E}_n) f_{n-1}(z)
dz,~(4.8A)\nonumber
\end{eqnarray}
and in Case $(B)$
\begin{eqnarray}\label{e4.8B}
f_n(r)=1 - 2 \int_{0}^{r}y^{-N+1}\phi^{-2}(y)dy
\int_{0}^{y}z^{N-1}\phi^{2}(z) (h(z)- {\cal E}_n) f_{n-1}(z)
dz.~~~(4.8B)\nonumber
\end{eqnarray}
\setcounter{equation}{8}In both cases, ${\cal E}_n$ is determined
by the corresponding $f_{n-1}(r)$ through (\ref{e2.22}); i.e.,
\begin{eqnarray}\label{e4.9}
{\cal E}_n=[h~f_{n-1}]/[f_{n-1}]
\end{eqnarray}
in which $[F]$ of any function $F(r)$ is defined to be
\begin{eqnarray}\label{e4.10}
[F]=\int_0^{\infty}r^{N-1}\phi^2(r) F(r) dr.
\end{eqnarray}
It is convenient to introduce $\chi(r)$ defined by
\begin{eqnarray}\label{e4.11}
\chi(r)\equiv r^{(N-1)/2}\phi(r).
\end{eqnarray}
Thus, (\ref{e4.10}) becomes
\begin{eqnarray}\label{e4.12}
[F]=\int_0^{\infty}\chi^2(r) F(r) dr.
\end{eqnarray}
Let $\rho_n(r)$, ${\cal K}(r)$ and ${\cal D}_n(r)$ be related to
$\sigma_n(r)$, $\kappa(r)$ and $(D_n)_r$ of (\ref{e3.2}),
(\ref{e3.20}) and (\ref{e3.22}) by
\begin{eqnarray}\label{e4.13}
\rho_n(r)=r^{N-1}\sigma_n(r)= (h(r)-{\cal E}_n)\chi^2(r)
f_{n-1}(r)
\end{eqnarray}
\begin{eqnarray}\label{e4.14}
{\cal K}(r)=r^{N-1}\kappa(r)=\chi^2(r)
\end{eqnarray}
and
\begin{eqnarray}\label{e4.15}
{\cal D}_n(r)=r^{N-1}(D_n)_r=-\frac{1}{2}{\cal K}(r)f'_n(r).
\end{eqnarray}
Correspondingly, in Case $(A)$, (4.8A) can be written as
\begin{eqnarray}\label{e4.16A}
~~~~~f_n(r)=1 - 2 \int_{r}^{\infty}\chi^{-2}(y)dy
\int_{y}^{\infty}\chi^{2}(z) (h(z)- {\cal E}_n) f_{n-1}(z)
dz,~~~~~~~(4.16A)\nonumber
\end{eqnarray}
and in Case $(B)$, (4.8B) becomes
\begin{eqnarray}\label{e4.16B}
~~~~~f_n(r)=1 - 2 \int_{0}^{r}\chi^{-2}(y)dy
\int_{0}^{y}\chi^{2}(z) (h(z)- {\cal E}_n) f_{n-1}(z)
dz.~~~~~~~~~(4.16B)\nonumber
\end{eqnarray}
\setcounter{equation}{16}

Because of (\ref{e3.23}), we have
\begin{eqnarray}\label{e4.17}
{\cal D}'_n(r)=\rho_n(r)
\end{eqnarray}
and therefore
\begin{eqnarray}\label{e4.18}
{\cal D}_n(r)=- \int_{r}^{\infty} \rho_n(z)dz = -
\int_{r}^{\infty} (h(z)- {\cal E}_n) \chi^2(z)f_{n-1}(z)dz.
\end{eqnarray}
From (\ref{e3.7}),
\begin{eqnarray}\label{e4.19}
\int_{0}^{\infty} \rho_n(r)dr=0
\end{eqnarray}
which leads to
\begin{eqnarray}\label{e4.20}
{\cal D}_n(r)= \int_0^{r} \rho_n(z)dz = \int_0^{r} (h(z)- {\cal
E}_n) \chi^2(z)f_{n-1}(z)dz.
\end{eqnarray}
These two expressions of ${\cal D}_n(r)$, (\ref{e4.18}) and
(\ref{e4.20}), are valid for both cases $(A)$ and $(B)$. Let $r_n$
be defined by
\begin{eqnarray}\label{e4.21}
h(r)-{\cal E}_n=0 ~~~~~~~{\sf at}~~~r=r_n.
\end{eqnarray}
Since $h'(r)<0$, (\ref{e4.21}) has one and only one solution, with
$h(r)-{\cal E}_n$ negative for $r>r_n$ and positive for $r<r_n$.
Thus, if
\begin{eqnarray}\label{e4.22}
f_{n-1}(r)>0
\end{eqnarray}
for all $r>0$, we have from (\ref{e4.18}) and (\ref{e4.20})
\begin{eqnarray}\label{e4.23}
{\cal D}_n(r)>0
\end{eqnarray}
and therefore, on account of (\ref{e4.15}),
\begin{eqnarray}\label{e4.24}
f'_n(r)<0.
\end{eqnarray}

In Case $(A)$, because of $f_n(\infty)=1$, (\ref{e4.24}) leads to
\begin{eqnarray}\label{e4.25A}
~~~~~~~~~~~~~~~~~~~~~~~~~~f_n(0)>f_n(r)>f_n(\infty)=1.~~~~~~~~~~~~~
~~~~~~~~~~~~(4.25A)\nonumber
\end{eqnarray}
Since for $n=0$, $f_0(r)=1$, (\ref{e4.22}) - (4.25A) are valid for
$n=1$; by induction these expressions also hold for all $n$; in
Case $(A)$, their validity imposes no restriction on the magnitude
of $h(r)$. In Case $(B)$ we assume $h(r)$ to be not too large, so
that (4.16B) is consistent with
\begin{eqnarray}
f_n(r) > 0~~~~~~~~{\sf for~all}~~r \nonumber
\end{eqnarray}
and therefore
\begin{eqnarray}\label{e4.25B}
~~~~~~~~~~~~~~~~~~~~~~~f_n(0)=1>f_n(r)>f_n(\infty)>0.~~~~~~~~~~
~~~~~~~~~~~~~(4.25B)\nonumber
\end{eqnarray}
\setcounter{equation}{25} As we shall see, these two boundary
conditions $(A)$ and $(B)$ produce sequences that have very
different behavior. Yet, they also share a number of common
properties.


\underline{Hierarchy Theorem} $(A)$ With the boundary condition
$f_n(\infty)=1$, we have for all $n$
\begin{eqnarray}\label{e4.26}
{\cal E}_{n+1} > {\cal E}_n
\end{eqnarray}
and
\begin{eqnarray}\label{e4.27}
\frac{d}{dr}\left( \frac{f_{n+1}(r)}{f_n(r)}\right)<0 ~~~~~{\sf
at~~any}~~~r>0.
\end{eqnarray}
Thus, the sequences $\{{\cal E}_n\}$ and $\{f_n(r)\}$ are all
monotonic, with
\begin{eqnarray}\label{e4.28}
{\cal E}_1<{\cal E}_2<{\cal E}_3<\cdots
\end{eqnarray}
and
\begin{eqnarray}\label{e4.29}
1<f_1(r)<f_2(r)<f_3(r)<\cdots
\end{eqnarray}
at all finite $r$.

$(B)$ With the boundary condition $f_n(0)=1$, we have for all odd
$n=2m+1$ an ascending sequence
\begin{eqnarray}\label{e4.30}
{\cal E}_1<{\cal E}_3<{\cal E}_5<\cdots,
\end{eqnarray}
but for all even $n=2m$, a descending sequence
\begin{eqnarray}\label{e4.31}
{\cal E}_2>{\cal E}_4>{\cal E}_6>\cdots;
\end{eqnarray}
furthermore, between any even $n=2m$ and any odd $n=2l+1$
\begin{eqnarray}\label{e4.32}
{\cal E}_{2m}>{\cal E}_{2l+1}.
\end{eqnarray}
Likewise, at any $r$, for any even $n=2m$
\begin{eqnarray}\label{e4.33}
\frac{d}{dr}\left( \frac{f_{2m+1}(r)}{f_{2m}(r)}\right)<0,
\end{eqnarray}
whereas for any odd $n=2l+1$
\begin{eqnarray}\label{e4.34}
\frac{d}{dr}\left( \frac{f_{2l+2}(r)}{f_{2l+1}(r)}\right)>0.
\end{eqnarray}

The proof of the hierarchy theorem is given in Appendix B. The
hierarchy theorem was discussed in our earlier papers for
one-dimensional problems[18,19]. The extension to  the
$N$-dimensional radially symmetric case is new. As we shall also
discuss in Appendix B, the lowest eigenvalue $E$ of the
Hamiltonian $T+V$ is the limit of the sequence $\{E_n\}$ with
\begin{eqnarray}\label{e4.35}
E_n=E_0-{\cal E}_n.
\end{eqnarray}
Thus, the boundary condition $f_n(\infty)=1$ yields a sequence, in
accordance with (\ref{e4.28}),
\begin{eqnarray}\label{e4.36}
E_1>E_2>E_3>\cdots>E,
\end{eqnarray}
with each member $E_n$ an {\it upper} bound of $E$, similar to the
usual variational method.

On the other hand, with the boundary condition $f_n(0) =1$, while
the sequence of its odd members $n=2l+1$ yields a similar one,
like (\ref{e4.36}), with
\begin{eqnarray}\label{e4.37}
E_1>E_3>E_5> \cdots >E,
\end{eqnarray}
its even members $n=2m$ satisfy
\begin{eqnarray}\label{e4.38}
E_2<E_4<E_6< \cdots <E.
\end{eqnarray}
It is unusual to have an iterative sequence of {\it lower} bounds
of the eigenvalue $E$. Together,  these sequences may be quite
efficient to pinpoint the limiting $E$.

In Appendix C, we discuss the comparison between the iterative
solution and perturbative series for some simple examples.

\newpage

\section*{\large \bf 5. Tribute}
\setcounter{section}{5} \setcounter{equation}{0}

This paper is based on my joint work with Richard Friedberg. It
was a pleasure for me to present the main body of the paper at the
Number and Nature Symposium held on December 17-18, 2004 at the
Rockefeller University, as a celebration for Mitchell J.
Feigenbaum's 60th birthday. In the Chinese tradition, sixty is
very special. It signifies the successful completion of one's
first cycle, and the fresh beginning of a new second cycle.

Because China has no indigenous religion, there is no origin of
reference in the Chinese calender, nothing equivalent to BC or AD.
All Chinese years are labelled cyclically, modulus sixty.

In Chinese calender, each year is designated by two indices. The
first index runs cyclically from one to ten, with each number
represented by a different word shown in the first column of
Chinese characters in Figure 1. We may translate the first
character at the top of that column as alpha (for one) and the
second character as beta (for two), then gamma (for three),
$\cdots$ until kappa (for ten); after kappa, we have alpha, beta,
$\cdots$ again. The second column of Chinese characters has twelve
words, each for a different animal. In Figure 1, the year 1944,
when Mitchell was born is designated (Alpha-Monkey), the next year
1945 is (Beta-Chicken), $\cdots$; ten years later, the first index
for 1954 becomes again alpha, but the second index is horse; two
years later for 1956, the first (number) index is gamma, but the
second (animal) index becomes monkey again. It takes
$\frac{1}{2}\cdot 10 \cdot 12 = 60$ years to complete the cycle.
This year 2004 is like 1944, the Alpha-Monkey year and Mitchell is
now in his new cycle, starting a new beginning.

the most beloved story in China is the book "Monkey" by Wu
Cheng-En written in the 16th century. It is about an immortal
Monkey-King of Chaos. The translation was by Arthur Waley (Figure
2) in 1943, perhaps anticipating that 1944 was the beginning of
Mitchell's Alpha-Monkey cycle. To celebrate 2004, the beginning of
Mitchell's newest cycle, Figure 3 is a drawing of mine, entitled
"Monkey King the Thinker". All of us are looking forward to this
new cycle with our very best wishes.

\newpage

\begin{center}
{\Large \bf \sf Figure Captions}
\end{center}

1. Chinese calender is cyclic, modulus sixty.\\

2. One of the most beloved novels in China\\

3. "Monkey King the Thinker" by T. D. Lee (2004) (color online)

\newpage

\centerline{\epsfig{file=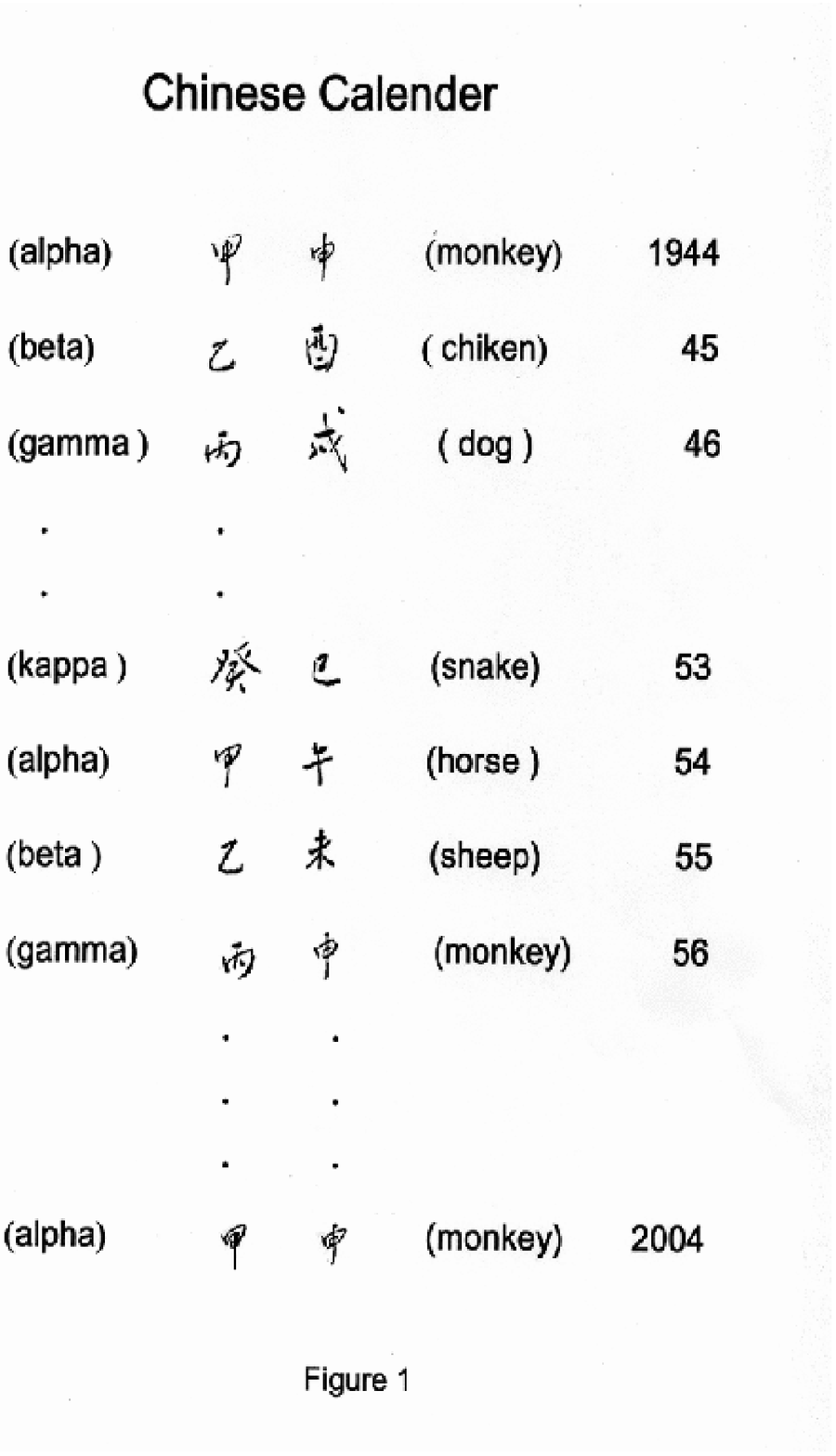,width=10cm}}
\centerline{\epsfig{file=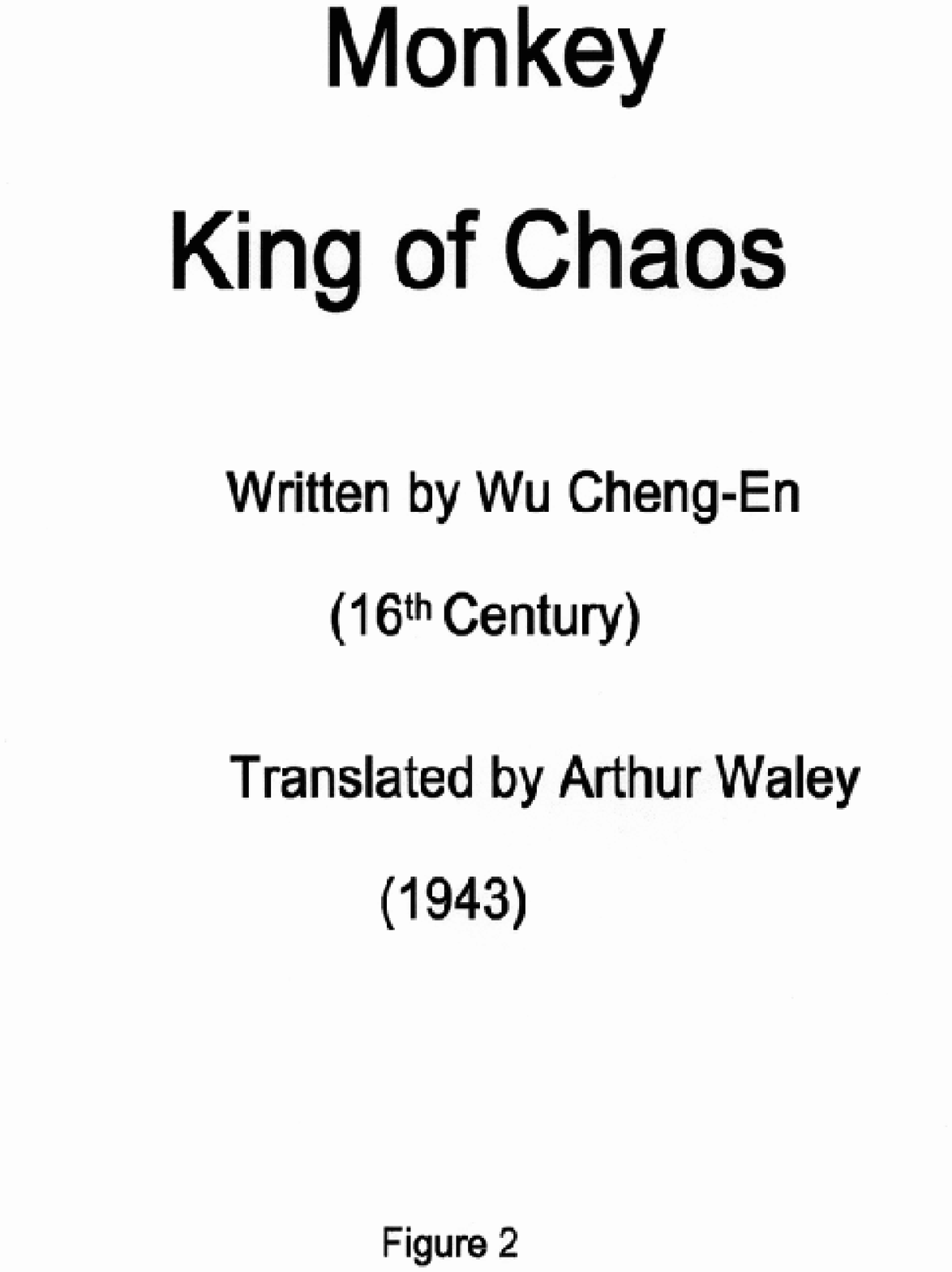,width=10cm}}
\centerline{\epsfig{file=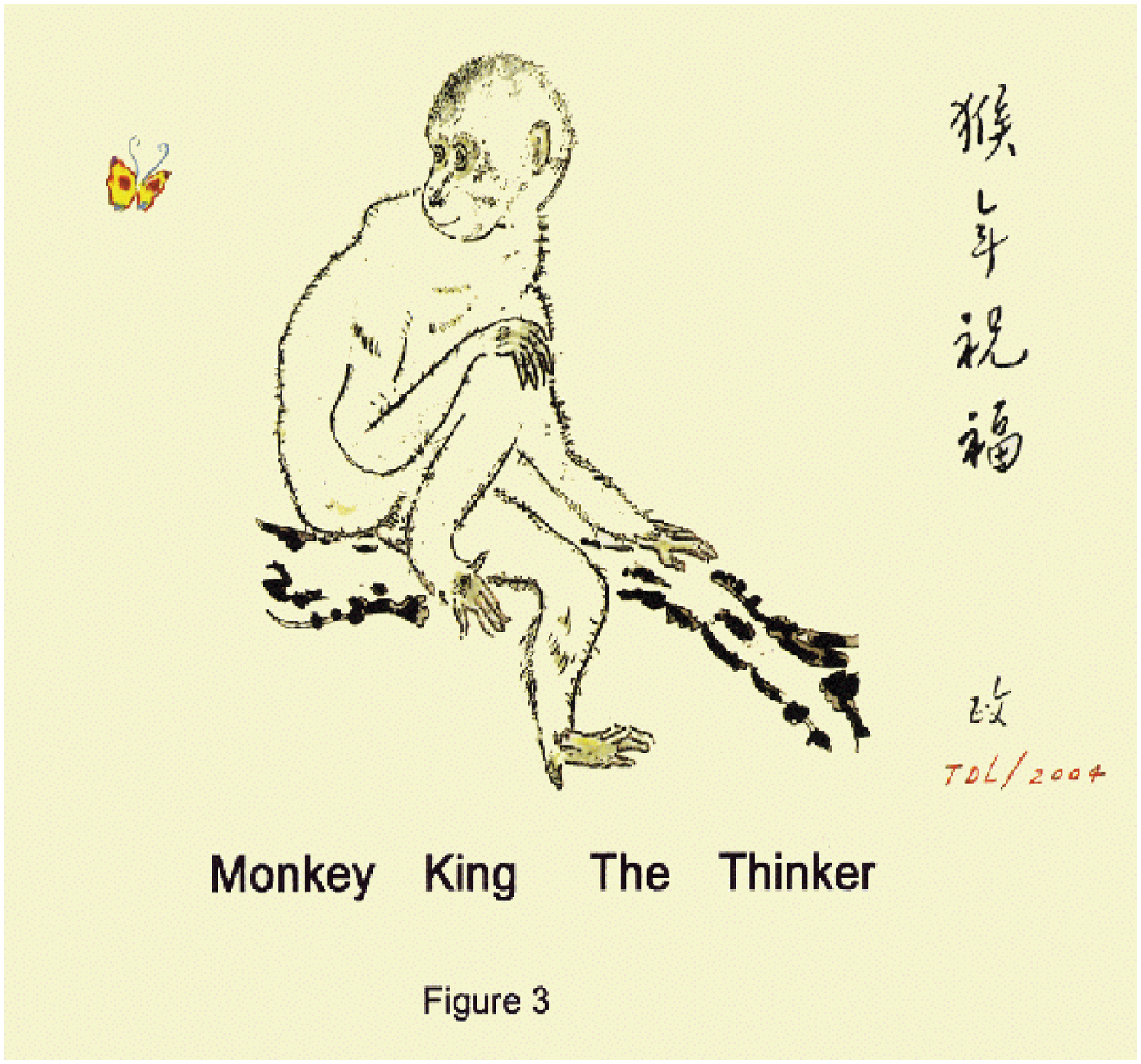,width=14cm}}

\newpage

\section*{\bf References}
\setcounter{section}{10}
\setcounter{equation}{0}

\noindent [1] W. Marciano, hep-ph/0411179

\noindent [2] M. Davier and W. Marciano, Annu. Rev. Nucl. Part.
Sci. 54(2004), 115

\noindent [3] M. Passera, hep-ph/0411168

\noindent [4] F. Dyson (unpublished)

\noindent
[5] A. M. Polyakov, Nucl.Phys. B121 (1977), 429 \\
~[6] G. 't Hooft, in: A. Zichichi, Erice(Eds.), The why's of
subnuclear physics

\noindent ~~~~~~~~~~        Plenum, New York, 1977\\
~[7] E. Brezin, G. Parisi and J. Zinn-Justin, Phys.Rev. D16 (1977), 408\\
~[8] J. Zinn-Justin, J.Math.Phys. 22 (1981), 511 \\
~[9] J. Zinn-Justin, Nucl.Phys. B192 (1981), 125 \\
~[10] J. Zinn-Justin, in: J.-D. Zuber, R. Stora (Eds.), Recent
advances in field

\noindent ~~~~~~~~~theory and statistical mechanics,
Les Houches, session XXXIX, 1982 \\
~[11] J. Zinn-Justin, Private Communication \\
~[12] Sidney Coleman, Aspects of Symmetry, Press Syndicate of

\noindent ~~~~~~~~~the University of Cambridge, 1987  \\
~[13] E. Shuryak, Nucl.Phys. B302 (1988), 621

\noindent [14] S. V. Faleev and P. G. Silvestrov, Phys. Lett. A197
(1995), 372

\noindent [15] R. Friedberg, T. D. Lee and W. Q. Zhao

\noindent ~~~~~~~~~ IL Nuovo Cimento A112 (1999), 1195

\noindent [16] R. Friedberg, T. D. Lee and W. Q. Zhao, Ann.Phys.
288 (2001), 52

\noindent
[17] R. Friedberg, T. D. Lee, W. Q. Zhao and A. Cimenser

\noindent ~~~~~~~~~Ann.Phys. 294 (2001), 67\\
\noindent [18] R. Friedberg and T. D. Lee, Ann.Phys. 308 (2003),
263\\
\noindent [19] R. Friedberg and T. D. Lee, Preprint quant-ph/0407207

\noindent ~~~~~~~~~Ann. Phys. (in press)\\
\noindent [20] See, e.g., Philip M. Morse and Herman Feshbach,

\noindent ~~~~~~~~~Methods of Theoretical Physics, McGraw-Hill
Brook Co. (1953)

\newpage

\section*{\bf Appendix A Construction of Trial Functions}
\setcounter{section}{7} \setcounter{equation}{0}

\noindent {\bf A.1 A New Formulation of Perturbative Expansion}\\

As mentioned in Section 1, in many problems of interest,
perturbative expansion lead to asymptotic series. Nevertheless,
the first few terms of such an expansion could provide important
insight to what a good trial function might be. For our purpose, a
particularly convenient way is to follow the method developed in
Refs.[15] and [16]. As we shall see, in  this new method to each
order of the perturbation, the wave function is always expressible
in terms of a single line-integral in the N-dimensional coordinate
space, which can be readily used for the construction of the trial
wave function.

We begin with the Hamiltonian $H$ in its standard form
(\ref{e2.1})-(\ref{e2.3}). Assume $V(q)$ to be positive
definite, and choose its minimum to be at $q=0$, with
$$
V(q) \geq V(0)=0.
\eqno(A.1)
$$
Introduce a scale factor $g^2$ by writing
$$
V(q) = g^2 v(q)
\eqno(A.2)
$$
and correspondingly
$$
\psi(q) = e^{-g S(q)}.
\eqno(A.3)
$$
Thus, the Schroedinger equation (\ref{e2.4}) becomes
$$
\bigg(-\frac{1}{2}\nabla^2 + g^2 v(q)\bigg)~e^{-g S(q)}=E e^{-g
S(q)}
\eqno(A.4)
$$
where, as before, $q$ denotes $q_1,~q_2,~\cdots,~q_N$ and
$\nabla$ the corresponding gradient operator. Hence $S(q)$
satisfies
$$
-\frac{1}{2}g^2(\nabla S)^2 + \frac{1}{2}g\nabla^2 S + g^2 v = E.
\eqno(A.5)
$$
Considering the case of large $g$, we expand
$$
S(q) = S_0(q) + g^{-1}S_1(q) + g^{-2}S_2(q) + \cdots
\eqno(A.6)
$$
and
$$
E = gE_0 + E_1 + g^{-1}E_2  + \cdots.
\eqno(A.7)
$$
Substituting (A.6) - (A.7) into (A.5) and
equating the coefficients of $g^{-n}$ on both sides, we find
\begin{eqnarray}\label{eA.8}
~~~~~~~~~~~~~~~(\nabla S_0)^2 &=& 2v, \nonumber\\
\nabla S_0 \cdot \nabla S_1 &=& \frac{1}{2}~\nabla^2 S_0
-E_0,\nonumber\\
\nabla S_0 \cdot \nabla S_2 &=& \frac{1}{2} ~[\nabla^2
S_1 - (\nabla S_1)^2] -E_1,~~~~~~~~~~~~~~~~~~~~~~~~~~~(A.8)\nonumber\\
\nabla S_0 \cdot \nabla S_3 &=& \frac{1}{2} ~[\nabla^2 S_2 - 2~
\nabla S_1 \cdot \nabla S_2] -E_2, \nonumber
\end{eqnarray}
etc. In this way, the second order partial differential equation
(A.5) is reduced to a series of first order partial
differential equations (A.8). The first of this set of
equations can be written as
$$
\frac{1}{2}[\nabla S_0(q)]^2 - v(q)=0+.
\eqno(A.9)
$$

As noted in Ref.[15], this is precisely the Hamilton-Jacobi
equation of a single particle with unit mass moving in a potential
"$-v(q)$" in the N-dimensional $q$-space. Since $q=0$ is the {\it
maximum} of the classical potential energy function $-v(q)$, for
any point $q \neq 0$ there is always a classical trajectory with a
total energy $0+$, which begins from $q=0$ and ends at the other
point $q \neq 0$, with $S_0(q)$ given by the corresponding
classical action integral. Furthermore, $S_0(q)$ increases along
the direction of the trajectory, which can be extended beyond the
selected point $q \neq 0$, towards $\infty$. At infinity, it is
easy to see that $S_0(q) = \infty$, and therefore the
corresponding wave amplitude $e^{-gS_0(q)}$ is zero. To solve the
second equation in (A.8), we note that, in accordance with (A.1) -
(A.2) at $q=0$, $\nabla S_0 \propto v^{\frac{1}{2}}(0)=0$. By
requiring $S_1(q)$ to be analytic at $q=0$, we determine
$$
E_0 = \frac{1}{2}(\nabla^2 S_0)_{{\sf at}~q= 0}.
\eqno(A.10)
$$
It is convenient to consider the surface
$$
S_0(q) = {\sf constant};
\eqno(A.11)
$$
its normal is along the corresponding classical trajectory passing
through $q$. Characterize  each classical trajectory by the
$S_0$-value along the trajectory and a set of $N-1$ angular
variables
$$
\alpha = (\alpha_1(q),\alpha_2(q), \cdots,
\alpha_{N-1}(q)),
\eqno(A.12)
$$
so that each $\alpha$ determines one classical trajectory with
$$
\nabla \alpha_j \cdot \nabla S_0 = 0,
\eqno(A.13)
$$
where
$$
j = 1, 2, \cdots, N-1.
\eqno(A.14)
$$
(As an example, we note that as $q \rightarrow 0$, $v(q)
\rightarrow \frac{1}{2} \sum\limits_i \omega_i^2 q_i^2$ and
therefore $ S_0 \rightarrow \frac{1}{2} \sum\limits_i \omega_i
q_i^2$. Consider  the ellipsoidal surface $ S_0 =$ constant. For
$S_0$ sufficiently small, each classical trajectory is normal to
this ellipsoidal surface. A convenient choice of $\alpha$ could be
simply any $N-1$ orthogonal parametric coordinates on the
surface.) Each $\alpha$ designates one classical trajectory, and
vice versa. Every $(S_0, \alpha)$ is mapped into a unique set
$(q_1,~q_2,~\cdots,~q_N)$ with $ S_0 \geq 0$ by construction. In
what follows, we regard the points in the $q$-space as specified
by the coordinates $( S_0, \alpha)$. Depending on the problem, the
mapping $(q_1,~q_2,~\cdots,~q_N) \rightarrow (S_0,\alpha)$ may or
may not be one-to-one. We note that, for $q$ near $0$, different
trajectories emanating  from $q=0$ have to go along different
directions, and therefore must associate with different $\alpha$.
Later on, as $S_0$ increases each different trajectory retains its
initially different $\alpha$-designation; consequently, using
$(S_0,\alpha)$ as the primary coordinates, different trajectories
never cross each other. The trouble-some complications of
trajectory-crossing in $q$-space is automatically resolved by
using $(S_0,\alpha)$ as coordinates. Keeping $\alpha$ fixed, the
set of first order partial differential equations can be further
reduced to a set of first order ordinary differential equations,
which are readily solvable, as we shall see.

Write
$$
S_1(q) = S_1(S_0, \alpha),
\eqno(A.15)
$$
the second line of (A.8) becomes
$$
(\nabla S_0)^2 (\frac{\partial S_1}{\partial S_0})_{\alpha} =
\frac{1}{2}\nabla^2 S_0 -E_0,
\eqno(A.16)
$$
and leads to, besides (A.10), also
$$
S_1(q) = S_1(S_0, \alpha) = \int\limits_0^{S_0} \frac{d
S_0}{(\nabla S_0)^2} [\frac{1}{2} \nabla^2 S_0 - E_0],
\eqno(A.17)
$$
where the integration is taken along the classical trajectory of
constant $\alpha$. Likewise, the third, fourth and other lines of
(A.8) lead to
$$
E_1 = \frac{1}{2} [ \nabla^2 S_1 - ( \nabla S_1)^2 ]_
{{\rm at}~q= 0},
\eqno(A.18)
$$
$$
 S_2(q) = S_2( S_0, \alpha) =
\int\limits_0^{S_0} \frac{d S_0}{(\nabla S_0)^2} \{\frac{1}{2}
[\nabla^2
S_1 - ( \nabla S_1)^2] - E_1\},
\eqno(A.19)
$$
$$
E_2 = \frac{1}{2} [\nabla^2 S_2 - 2 ( \nabla S_1)\cdot
( \nabla S_2) ]_{{\rm at}~q= 0},
\eqno(A.20)
$$
$$
S_3(q) =  S_3( S_0, \alpha) = \int\limits_0^{ S_0} \frac{d
S_0}{( \nabla S_0)^2} \{\frac{1}{2} [ \nabla^2 S_2 - 2 ( \nabla
S_1) \cdot ( \nabla S_2)] - E_2\},
\eqno(A.21)
$$
etc. These solutions give the convenient normalization convention
at $q = 0$,
$$
S(0) = 0
$$
and
$$
e^{-S(0)} = 1.
\eqno(A.22)
$$


\noindent
\underline{Remarks}\\

\noindent(i) As an example, consider an N-dimensional harmonic
oscillator with
$$
V(q) = \frac{g^2}{2}(q_1^2+q_2^2+\cdots +q_N^2).
\eqno(A.23)
$$
From (A.2), one sees that the Hamilton-Jacobi equation
(A.9) is for a particle moving in a potential given by
$$
-v(q) = -\frac{1}{2}(q_1^2+q_2^2+\cdots +q_N^2).
\eqno(A.24)
$$
Thus, for any point $q \neq 0$ the classical trajectory of
interest is simply a straight line connecting the origin and the
specific point, with the action
$$
S_0(q) = \frac{1}{2}(q_1^2+q_2^2+\cdots +q_N^2).
\eqno(A.25)
$$
The corresponding energy is, in accordance with (A.10),
$$
E_0 = \frac{N}{2}.
\eqno(A.26)
$$
By using (A.8), one can readily show that
$E_1=E_2=\cdots=0$ and $S_1=S_2=\cdots=0$. The result is the well
known exact answer with the groundstate wave function for the
Schroedinger equation (A.4) given by
$$
e^{-gS(q)} = exp~[-\frac{g}{2}(q_1^2+q_2^2+\cdots +q_N^2)]
\eqno(A.27)
$$
and the corresponding energy
$$
E = \frac{N}{2}g.
\eqno(A.28)
$$

\noindent(ii) From this example, it is clear that the above
expressions (A.6) - (A.8) are {\it not} the well-known WKB method.
The new formalism uses $-v(q)$ as the potential for the
Hamilton-Jacobi equation, and its "classical" trajectory carries a
$0+$ energy; consequently, unlike the WKB method, there is no
turning point along the classical trajectory, and the formalism is
applicable to arbitrary dimensions.\\


\noindent {\bf A.2 Trial Function for the Quantum Double-well
Potential}\\

To illustrate how to construct a trial function, consider the
quartic potential (\ref{e1.3}) in one dimension with degenerate
minima. Set $a=1$, we have
$$
V(x) = \frac{1}{2}g^2 (x^2-1)^2.
\eqno(A.29)
$$
The corresponding Schroedinger equation is
$$
(-\frac{1}{2}\frac{d^2}{dx^2} + \frac{1}{2}g^2
(x^2-1)^2)\psi(x)=E\psi(x)
\eqno(A.30)
$$
where, as before, $\psi(x)=e^{-gS(x)}$ is the groundstate wave
function and $E$ its energy. Using the expansions (A.6) -
(A.7) and following the steps (A.8), (A.10)
and (A.15) - (A.21), we find the well-known perturbative
series
$$
S_0(x) = \frac{1}{3}(x-1)^2(x+2),~~~~~S_1(x) = \ln
\frac{x+1}{2},~~~~~ S_2(x) =
\frac{3}{16}-\frac{x+2}{4(x+1)^2},~~~\cdots
\eqno(A.31)
$$
and
$$
E_0=1,~~~E_1=-\frac{1}{4},~~~E_2=-\frac{9}{64},~~~\cdots.
\eqno(A.32)
$$
Both expansions $S=S_0+g^{-1}S_1+g^{-2}S_2+\cdots$ and
$E=gE_0+E_1+g^{-1}E_2+\cdots$ are divergent, furthermore, at
$x=-1$ and for $n\geq 1$, each $S_n(x)$ is infinite. The
reflection $x\rightarrow -x$ gives a corresponding asymptotic
expansion $S_n(x)\rightarrow S_n(-x)$, in which each $S_n(-x)$ is
regular at $x=-1$, but singular at $x=+1$.

As noted in section 1, for $g$ large, the first few terms of the
perturbative series (with (A.31) for $x$ positive and the
corresponding expansion $S_n(x)\rightarrow S_n(-x)$ for $x$
negative) give a fairly good description of the true wave function
$\psi(x)$ whenever $\psi(x)$ is large (i.e. for $x$ near $\pm 1$).
However, for $x$ near zero, when $\psi(x)$ is exponentially small,
the perturbative series becomes totally unreliable. This suggests
the use of first few terms of the perturbative series for regions
whenever $\psi(x)$ is expected to be large. In regions where
$\psi(x)$ is exponentially small, simple interpolations by hand
may already be adequate for a trial function, as we shall see.
Since the quartic potential (A.29) is even in $x$, so is
the groundstate wave function; likewise, we require the trial
function $\phi(x)$ also to satisfy $\phi(x)=\phi(-x)$. At $x=0$,
we require
$$
(\frac{d\phi}{dx})_{x=0} =0. \eqno(A.33)
$$
To construct $\phi(x)$, we start with the first two functions
$S_0(x)$ and $S_1(x)$ in (A.31). Introduce, for $x\geq 0$,
$$
\phi_+(x) \equiv e^{-gS_0(x)-S_1(x)}=(\frac{2}{1+x})e^{-gS_0(x)}
\eqno(A.34)
$$
and
$$
\phi_-(x) \equiv
e^{-gS_0(-x)-S_1(x)}=(\frac{2}{1+x})e^{-\frac{4}{3}g+gS_0(x)}.
\eqno(A.35)
$$
In order to satisfy (A.33), we define
\begin{eqnarray}\label{eA.36}
\phi(x)=\phi(-x) \equiv \left\{
\begin{array}{ll}
\phi_+(x) + \frac{g-1}{g+1}\phi_-(x), &~~~~ {\sf for} ~~~0\leq x<1\\
(1+ \frac{g-1}{g+1}e^{-\frac{4}{3}g})\phi_+(x), &~~~~ {\sf for}
~~~x>1
\end{array}
\right.~~~~~~~~~~~(A.36)\nonumber
\end{eqnarray}
Thus, by construction (A.33) is satisfied. Furthermore, $\phi(x)$
is continuous everywhere, for $x$ from $-\infty$ to $\infty$, and
so is its derivative.

By differentiating $\phi_+(x)$ and $\phi(x)$, we see that they
satisfy
$$
(T+V+u)\phi_+=g\phi_+
\eqno(A.37)
$$
and
$$
(T+V+h)\phi=g\phi,
\eqno(A.38)
$$
where
$$
u(x)=\frac{1}{(1+x)^2}
\eqno(A.39)
$$
and
$$
h(x)=h(-x)
\eqno(A.40)
$$
with, for $x \geq 0$
$$
h(x)=u(x) + \hat{g} (x)
\eqno(A.41)
$$
where
\begin{eqnarray}\label{eA.42}
~~~~~~~~\hat{g} (x) = \left\{\begin{array}{ll} 2g\frac{(g-1) e^{2g
S_0(x)-\frac{4}{3} g}}{(g+1)+(g-1)e^{2g S_0(x)-\frac{4}{3} g}},
&~~~~~~{\sf for}~~0 \leq x<1\\
0&~~~~~~{\sf for}~~x>1.
\end{array}
\right.~~~~~~~~~~~~(A.42)\nonumber
\end{eqnarray}
Note that for $g>1$, $\hat{g}(x)$ is positive, and has a
discontinuity at $x=1$. Furthermore, for $x$ positive both $u(x)$
and $\hat{g}(x)$ are decreasing functions of $x$. Therefore,
$h(x)$ also satisfies (\ref{e4.4}) for $x=r>0$.

\newpage

\section*{\bf Appendix B.~~~ Proof of the Hierarchy Theorem }
\setcounter{section}{8} \setcounter{equation}{0}

In this Appendix, we give the proof of the Hierarchy Theorem
stated in Section~4. The discussion follows very closely the one
given for the one-dimensional case in Ref. 19. We first establish
several lemmas applicable to {\it both} boundary conditions
(\ref{e4.6}) and (\ref{e4.7}): $(A)$ $f_n(\infty)=1$ and $(B)$
$f_n(0)=1$.

\noindent \underline{Lemma 1} ~~For any pair $f_m(r)$ and $f_l(r)$
$$
{\sf i)~~if~~at~all}~r,~~~\frac{d}{dr}\left(
\frac{f_m(r)}{f_l(r)}\right) <0~~~~~{\sf then}~~~{\cal
E}_{m+1}>{\cal E}_{l+1}, \eqno(B.1)
$$
and
$$
{\sf ii)~~if~~at~all}~r,~~~\frac{d}{dr}\left(
\frac{f_m(r)}{f_l(r)}\right)
>0~~~~~{\sf then}~~~{\cal E}_{m+1}<{\cal E}_{l+1}.
\eqno(B.2)
$$
\underline{Proof}

According to (\ref{e4.9})
$$
{\cal E}_{m+1}[f_m]=[h~f_m].
\eqno(B.3)
$$
Also by definition (\ref{e4.10}),
$$
{\cal E}_{l+1}[f_m]=[{\cal E}_{l+1}~f_m].
\eqno(B.4)
$$
Their difference gives
$$
({\cal E}_{m+1}-{\cal E}_{l+1})[f_m]=[(h-{\cal E}_{l+1})~f_m].
\eqno(B.5)
$$
From (B.3),
$$
0=[(h-{\cal E}_{l+1})~f_l].
\eqno(B.6)
$$
Let $r_{l+1}$ be defined by (\ref{e4.21}); i.e., at $r=r_{l+1}$,
$$
h(r_{l+1})={\cal E}_{l+1}.
\eqno(B.7)
$$
Multiplying (B.5) by $f_l(r_{l+1})$ and (B.6) by
$f_m(r_{l+1})$ and taking their difference, we have
$$
f_l(r_{l+1})({\cal E}_{m+1}-{\cal E}_{l+1})[f_m]=[(h-{\cal
E}_{l+1})~(f_m(r)f_l(r_{l+1})-f_l(r)f_m(r_{l+1}))],
\eqno(B.8)
$$
in which the unsubscripted $r$ acts as a dummy variable; thus
$[f_m(r)]$ means $[f_m]$ and $[f_m(r_{l+1})]$ means
$f_m(r_{l+1})\cdot [\hspace*{.04cm}1\hspace*{.04cm}]$, etc.

(i) If $(f_m(r)/f_l(r))'<0$, then for $r<r_{l+1}$
$$
\frac{f_m(r)}{f_l(r)} > \frac{f_m(r_{l+1})}{f_l(r_{l+1})}.
\eqno(B.9)
$$
In addition, since in accordance with (\ref{e4.4}) and (B.7),
$h'(r)<0$ and $h(r_{l+1})={\cal E}_{l+1}$, we also have for
$r<r_{l+1}$
$$
h(r)>{\cal E}_{l+1}.
\eqno(B.10)
$$
Thus,  the function inside  the square bracket on the right hand
side of (B.8) is positive for $r<r_{l+1}$. Also, the
inequalities (B.9) and (B.10) both reverse their
signs for $r>r_{l+1}$. Consequently, the right hand side of
(B.8) is positive definite, and so is its left side.
Therefore, on account of (4.25A)-(4.25B), (B.1) holds.

(ii) If $(f_m(r)/f_l(r))'>0$, we see   that for $r<r_{l+1}$,
(B.9) reverses its sign but not (B.10). A similar
reversal of sign happens for $r>r_{l+1}$. Thus, the right hand
side of (B.8) is now negative definite and therefore
${\cal E}_{m+1}<{\cal E}_{l+1}$. Lemma 1 is proved.

Let
$$
\eta = \eta(\xi)
\eqno(B.11)
$$
be a single valued differentiable function of $\xi$ in the range
between $a$ and $b$ with
$$
0 \leq a \leq \xi \leq b
\eqno(B.12)
$$
and with
$$
\eta(a) \geq 0.
\eqno(B.13)
$$

\noindent \underline{Lemma 2}.

(i) The ratio $\eta/\xi$ is a decreasing function of $\xi$ for
$a<\xi<b$ if
$$
\frac{d\eta}{d\xi} \leq \frac{\eta}{\xi}~~~~~{\sf at}~~~\xi=a
\eqno(B.14)
$$
and
$$
\frac{d^2\eta}{d\xi^2} < 0~~~~~~{\sf for}~~~a<\xi<b.
\eqno(B.15)
$$

(ii) The ratio $\eta/\xi$ is an increasing function of $\xi$ for
$a<\xi<b$ if
$$
\frac{d\eta}{d\xi} \geq \frac{\eta}{\xi}~~~~~{\sf at}~~~\xi=a
\eqno(B.16)
$$
and
$$
\frac{d^2\eta}{d\xi^2} > 0~~~~~~{\sf for}~~~a<\xi<b.
\eqno(B.17)
$$
\underline{Proof} ~~Define
$$
L \equiv \xi \frac{d\eta}{d\xi} - \eta
\eqno(B.18)
$$
to be the Legendre transform $L(\xi)$. We have
$$
\frac{dL}{d\xi} = \xi\frac{d^2\eta}{d\xi^2}
\eqno(B.19)
$$
and
$$
\frac{d}{d\xi}\left(\frac{\eta}{\xi}\right) = \frac{L}{\xi^2}.
\eqno(B.20)
$$
Since (B.14) says that $L(a) \leq 0$ and (B.15)
says that $\frac{dL}{d\xi}<0$  for $a<\xi<b$, these two conditions
imply $L(\xi)<0$ for $a<\xi<b$, which proves (i) in view of
(B.20). The proof of (ii) is the same, but with
inequalities reversed.\\


\noindent \underline{Lemma 3} ~~For any pair $f_m(r)$ and $f_l(r)$

(i) if over all $r$,
$$
\frac{d}{dr}\left(\frac{f_m(r)}{f_l(r)}\right)<0~~~~ {\sf
then~~at~all}~r,~~~~ \frac{d}{dr}\left(\frac{{\cal
D}_{m+1}(r)}{{\cal D}_{l+1}(r)}\right)<0, \eqno(B.21)
$$
and (ii) if over all $r$,
$$
\frac{d}{dr}\left(\frac{f_m(r)}{f_l(r)}\right)>0~~~~{\sf
then~~at~all}~r,~~~~ \frac{d}{dr}\left(\frac{{\cal
D}_{m+1}(r)}{{\cal D}_{l+1}(r)}\right)>0. \eqno(B.22)
$$
\underline{Proof} ~~From (\ref{e4.18}) and (\ref{e4.20}), we have
$$
{\cal D}'_{m+1}(r)=(h(r)-{\cal E}_{m+1})~\chi^2(r)~f_m(r)
\eqno(B.23)
$$
and
$$
{\cal D}'_{l+1}(r)=(h(r)-{\cal E}_{l+1})~\chi^2(r)~f_l(r).
\eqno(B.24)
$$
Define
$$
\xi={\cal D}_{l+1}(r)~~{\sf and}~~\eta={\cal D}_{m+1}(r).
\eqno(B.25)
$$
In any local region of $r$ where ${\cal D}'_{l+1}(r) \neq 0$, we can
regard $\eta = \eta(\xi)$ through $\eta(r) = \eta(\xi(r))$. Hence,
we have
$$
\frac{d\eta}{d\xi}=\frac{{\cal D}'_{m+1}(r)}{{\cal D}'_{l+1}(r)}=R(r)
\frac{f_{m}(r)}{f_{l}(r)}
\eqno(B.26)
$$
where
$$
R(r)=\frac{h(r)-{\cal E}_{m+1}}{h(r)-{\cal E}_{l+1}},
\eqno(B.27)
$$
and
\begin{eqnarray}\label{eB.28}
~~~~~~~~~~~~~~~\frac{d}{d\xi}\left(\frac{d\eta}{d\xi}\right) &=&
\frac{1}{{\cal D}'_{l+1}} \left(\frac{{\cal D}'_{m+1}}{{\cal D}'_{l+1}}\right)'=
\frac{1}{{\cal D}'_{l+1}}\left( R \frac{f_{m}}{f_{l}}\right)'\nonumber\\
&=&\frac{1}{{\cal D}'_{l+1}}\left(R' \frac{f_{m}}{f_{l}}+R\left(
\frac{f_{m}}{f_{l}}\right)'\right)~~~~~~~~~~~~~~~~~~~~~~~(B.28)\nonumber
\end{eqnarray}
where
$$
R'(r)= \frac{{\cal E}_{m+1}-{\cal E}_{l+1}}{(h(r)-{\cal
E}_{l+1})^2}h'(r).
\eqno(B.29)
$$
(i) If $(f_m/f_l)'<0$, from Lemma 1, we have
$$
{\cal E}_{m+1}>{\cal E}_{l+1}.
\eqno(B.30)
$$
From $h'(r)<0$ and the definition of $r_{m+1}$ and $r_{l+1}$,
given by (B.7), we have
$$
r_{m+1}<r_{l+1},
\eqno(B.31)
$$
$$
h(r_{m+1})={\cal E}_{m+1}~~~~~{\sf and}~~~~~h(r_{l+1})={\cal
E}_{l+1}.
\eqno(B.32)
$$

We note that from (\ref{e4.18}) and (\ref{e4.20}) ${\cal
D}_{m+1}(r)$ and ${\cal D}_{l+1}(r)$ are both positive continuous
functions of $r$, varying from at $r=0$,
$$
{\cal D}_{m+1}(0)={\cal D}_{l+1}(0)=0
\eqno(B.33)
$$
to at $r=\infty$
$$
{\cal D}_{m+1}(\infty)={\cal D}_{l+1}(\infty)=0
\eqno(B.34)
$$
with their maxima at $r_{m+1}$ for ${\cal D}_{m+1}(r)$ and $r_{l+1}$ for
${\cal D}_{l+1}(r)$, since in accordance with (B.23)-(B.24)
and (B.32),
$$
{\cal D}'_{m+1}(r_{m+1})=0~~~~{\sf and}~~~~{\cal D}'_{l+1}(r_{l+1})=0.
\eqno(B.35)
$$
From (B.29)-(B.30), we see that $R'(r)$ is always
$<0$. Furthermore, from (B.27), we also find that the
function $R(r)$ has a discontinuity at $r=r_{l+1}$. At $r=0$,
$R(0)$ satisfies
$$
0<R(0)=\frac{h(0)-{\cal E}_{m+1}}{h(0)-{\cal E}_{l+1}}<1.
\eqno(B.36)
$$
As $r$ increases from $0$, $R(r)$ decreases from $R(0)$, through
$$
R(r_{m+1})=0,
\eqno(B.37)
$$
to $-\infty$ at $r=r_{l+1}-$; $R(r)$ then switches to $+\infty$ at
$r=r_{l+1}+$,  and continues to decrease as $r$ increases from
$r_{l+1}+$. At $r=\infty$, $R(r)$ becomes
$$
R(\infty)=\frac{{\cal E}_{m+1}}{{\cal E}_{l+1}}>1.
\eqno(B.38)
$$

It is convenient to divide the $r$-axis into three
regions:
\begin{eqnarray*}
~~~~~~~~~~~~~~~~~~~~~~~~~~~~~~~~
{\rm (I)}~~~&~ 0 < r < r_{m+1},~~~~~~~~~~~~~~~~~~~~~~~~~~~~~~~~ \\
{\rm (II)}~~&~r_{m+1} < r < r_{l+1}~~~~~~~~~~~~~~~~~~~~~~~~~~(B.39) \\
{\rm (III)}~&~ r_{l+1} < r.~~~~~~~~~~~~~~~~~~~~~~~~~~~~~~~~~
\end{eqnarray*}
The signs of ${\cal D}'_{m+1}$, ${\cal D}'_{l+1}$, $R$ and $R'$ in
these regions are summarized in Table~1. Assuming $(f_m/f_l)'<0$
we shall show the validity of (B.21), $({\cal D}_{m+1}/{\cal
D}_{l+1})'<0$, in each of these three regions.

\vspace{1cm}

\begin{center}
Table 1. The signs of ${\cal D}'_{m+1}(r)$, ${\cal D}'_{l+1}(r)$,
$h(r)-{\cal
E}_{m+1}$, $h(r)-{\cal E}_{l+1}$, \\
$R(r)$ and $R'(r)$ in the three regions defined by (B.39), when
${\cal E}_{m+1} >  {\cal E}_{l+1}$.
\end{center}

\begin{tabular}{|c|c|c|c|c|c|c|}
\hline
region& ${\cal D}'_{m+1}(r)$ & ${\cal D}'_{l+1}(r)$ & $h(r)-{\cal E}_{m+1}$ & $h(r)-{\cal E}_{l+1}$ &~$R(r)$~&~$R'(r)$~\\
\hline
{\rm I} & $>0$    & $>0$          &  $>0$             &  $>0$                 &  $>0$  & $<0$\\
{\rm II} & $<0$    & $>0$          &  $<0$             &  $>0$                 &  $<0$  & $<0$\\
{\rm III} & $<0$    & $<0$          &  $<0$             &  $<0$                 &  $>0$  & $<0$\\
\hline
\end{tabular}

\newpage

Since
$$
{\cal E}_{l+1}<h(r)<{\cal E}_{m+1}~~~~~{\sf in}~~~{\rm II},
\eqno(B.40)
$$
${\cal D}_{m+1}(r)$ is decreasing and ${\cal D}_{l+1}(r)$ is increasing; it is
clear that (B.21) holds in {\rm II}.

In each of regions ({\rm I}) and ({\rm III}), we have $R(r)>0$
from (B.27) and $R'(r)<0$ from (B.29). Since
$(f_m/f_l)'$ is always negative by the assumption in
(B.21), both terms inside the big parenthesis of
(B.28) are negative; hence the same (B.28) states
that $d^2\eta/d\xi^2$ has the opposite sign from ${\cal D}'_{l+1}$. From
the sign of ${\cal D}'_{l+1}$ listed in Table 1, we see that
$$
\frac{d^2\eta}{d\xi^2}<0~~~~~~~{\sf in}~~~({\rm I})
\eqno(B.41)
$$
and
$$
\frac{d^2\eta}{d\xi^2}>0~~~~~~~{\sf in}~~~({\rm III}).
\eqno(B.42)
$$
Within each region, $\eta={\cal D}_{m+1}(r)$ and $\xi={\cal D}_{l+1}(r)$ are
both monotonic in $r$; therefore, $\eta$ is a single-valued
function of $\xi$ and we can apply Lemma 2. In ({\rm I}), at
$r=0$, both ${\cal D}_{m+1}(0)$ and ${\cal D}_{l+1}(0)$ are $0$ according to
(\ref{e4.20}), but their ratio is given by
$$
\frac{{\cal D}_{m+1}(0)}{{\cal D}_{l+1}(0)}=\frac{{\cal D}'_{m+1}(0)}{{\cal D}'_{l+1}(0)}.
\eqno(B.43)
$$
Therefore,
$$
\left(\frac{d\eta}{d\xi}\right)_{r=0}=\left(\frac{\eta}{\xi}\right)_{r=0}.
\eqno(B.44)
$$
Furthermore, from (B.41), $\frac{d^2\eta}{d\xi^2}<0$ in
({\rm I}), it follows from Lemma 2, case(i), the ratio $\eta/\xi$
is a decreasing function of $\xi$. Since $\xi'={\cal D}'_{l+1}$ is $>0$
in ({\rm I}) according to (B.24), we have
$$
\frac{d}{dr}\left(\frac{{\cal D}_{m+1}}{{\cal D}_{l+1}}\right)<0~~~~~{\sf
in}~~~({\rm I}).
\eqno(B.45)
$$

In ({\rm III}), at $r=\infty$,   both ${\cal D}_{m+1}(\infty)$ and
${\cal D}_{l+1}(\infty)$ are $0$ according to (B.34). Their ratio
is
$$
\frac{{\cal D}_{m+1}(\infty)}{{\cal D}_{l+1}(\infty)}=
\frac{{\cal D}'_{m+1}(\infty)}{{\cal D}'_{l+1}(\infty)},
\eqno(B.46)
$$
which gives at $r=\infty$,
$$
\left(\frac{d\eta}{d\xi}\right)_{r=\infty}=\left(\frac{\eta}{\xi}\right)_{r=\infty}.
\eqno(B.47)
$$
As $r$ decreases from $r=\infty$ to $r=r_{l+1}$, from (B.42) we
have $\frac{d^2\eta}{d\xi^2}>0$ in ({\rm III}). It follows from
Lemma 2, case (ii), $\eta/\xi$ is an increasing function of $\xi$.
Since $\xi'={\cal D}'_{l+1}$ is $<0$, because $r>r_{l+1}$, we have
$$
\frac{d}{dr}\left(\frac{{\cal D}_{m+1}}{{\cal
D}_{l+1}}\right)<0~~~~~{\sf in}~~~({\rm III}). \eqno(B.48)
$$
Thus, we prove case(i) of Lemma 3. Case(ii) of Lemma 3 follows
from case (i) through the interchange of the subscripts $m$ and
$l$. Lemma 3 is  then established.

\noindent \underline{Lemma 4} ~~Take any pair $f_m(r)$ and
$f_l(r)$

\noindent $(A)$ For the boundary condition $f_n(\infty)=1$, if at
all $r$,
$$
\frac{d}{dr}\left(\frac{f_m(r)}{f_l(r)}\right)<0~~{\sf
then~~at~all}~r,~~
\frac{d}{dr}\left(\frac{f_{m+1}(r)}{f_{l+1}(r)}\right)<0;
\eqno(B.49A)
$$
therefore, if at all $r$,
$$
\frac{d}{dr}\left(\frac{f_m(r)}{f_l(r)}\right)>0~~{\sf
then~~at~all}~r,~~
\frac{d}{dr}\left(\frac{f_{m+1}(r)}{f_{l+1}(r)}\right)>0.
\eqno(B.50A)
$$
$(B)$ For the boundary condition $f_n(0)=1$, if at all $r$,
$$
\frac{d}{dr}\left(\frac{f_m(r)}{f_l(r)}\right)<0~~{\sf
then~~at~all}~r,~~
\frac{d}{dr}\left(\frac{f_{m+1}(r)}{f_{l+1}(r)}\right)>0;
\eqno(B.49B)
$$
therefore, if at all $r$,
$$
\frac{d}{dr}\left(\frac{f_m(r)}{f_l(r)}\right)>0~~{\sf
then~~at~all}~r,~~
\frac{d}{dr}\left(\frac{f_{m+1}(r)}{f_{l+1}(r)}\right)<0.
\eqno(B.50B)
$$
\underline{Proof}~~Define
$$
\hat{\xi} = f_{l+1}(r)~~~~~~~{\sf and}~~~~~~\hat{\eta}
=f_{m+1}(r).
\eqno(B.51)
$$
From (\ref{e4.15}) we see that
$$
\frac{d\hat{\eta}}{d\hat{\xi}} = \frac{f'_{m+1}(r)}{f'_{l+1}(r)} =
\frac{{\cal D}_{m+1}(r)}{{\cal D}_{l+1}(r)}
\eqno(B.52)
$$
and
$$
\frac{d}{d\hat{\xi}}\left( \frac{d\hat{\eta}}{d\hat{\xi}}\right)
=\frac{1}{f'_{l+1}}\frac{d}{dr}\left(
\frac{{\cal D}_{m+1}(r)}{{\cal D}_{l+1}(r)}\right).
\eqno(B.53)
$$
$(A)$ In this case $f_n(\infty)=1$  for all $n$. Thus, at
$r=\infty$, $\hat{\xi}=f_{l+1}(\infty)=1$,
$\hat{\eta}=f_{m+1}(\infty)=1$, and their ratio
$$
\left(\frac{\hat{\eta}}{\hat{\xi}}\right)_{r=\infty} = 1.
\eqno(B.54)
$$
At the same point $r=\infty$, in accordance with (\ref{e4.18}),
${\cal D}_{l+1}(\infty)={\cal D}_{m+1}(\infty)=0$, but their ratio is, on
account of $h(\infty)=0$ and (B.1) of Lemma 1,
$$\label{e2.80}
\frac{{\cal D}_{m+1}(\infty)}{{\cal D}_{l+1}(\infty)}=
\frac{{\cal D}'_{m+1}(\infty)}{{\cal D}'_{l+1}(\infty)}=
\frac{h(\infty)-{\cal E}_{m+1}}{h(\infty)-{\cal
E}_{l+1}}=\frac{{\cal E}_{m+1}}{{\cal E}_{l+1}}>1,
\eqno(B.55)
$$
in which the last inequality follows from the same assumption, if
$(f_m/f_l)'<0$, shared by (B.1) of Lemma 1 and the
present (B.49A) that we wish to prove. Thus, from (B.52),
at $r=\infty$
$$
\left(\frac{d\hat{\eta}}{d\hat{\xi}}\right)_{r=\infty}=
\frac{{\cal E}_{m+1}}{{\cal E}_{l+1}} >
\left(\frac{\hat{\eta}}{\hat{\xi}}\right)_{r=\infty}=1.
\eqno(B.56)
$$
As $r$ decreases from $\infty$ to $0$, $\hat{\xi}$ increases from
$f_{l+1}(\infty)=1$ to $f_{l+1}(0)>1$, in accordance with
(\ref{e4.24}) and (4.25A). On account of (B.21) of Lemma 3,
we have $({\cal D}_{m+1}/{\cal D}_{l+1})'<0$, which when combined with
(B.53) and $f'_n(r)<0$ leads to
$$
\frac{d}{d\hat{\xi}}\left(\frac{d\hat{\eta}}{d\hat{\xi}}\right)>0.
\eqno(B.57)
$$
Thus, by using (B.16)-(B.17) of Lemma 2, we have
$\hat{\eta}/\hat{\xi}$ to be an increasing function of
$\hat{\xi}$; i.e.,
$$
\frac{d}{d\hat{\xi}}\left(\frac{\hat{\eta}}{\hat{\xi}}\right)>0.
\eqno(B.58)
$$
Because
$$
\frac{d}{dr}\left(\frac{\hat{\eta}}{\hat{\xi}}\right)=
\hat{\xi}'\frac{d}{d\hat{\xi}}\left(\frac{\hat{\eta}}{\hat{\xi}}\right)=
f'_{l+1}\frac{d}{d\hat{\xi}}\left(\frac{\hat{\eta}}{\hat{\xi}}\right)
\eqno(B.59)
$$
and $f'_{l+1}<0$, we find
$$
\frac{d}{dr}\left( \frac{f_{m+1}}{f_{l+1}}\right)=
\frac{d}{dr}\left(\frac{\hat{\eta}}{\hat{\xi}}\right)<0,
\eqno(B.60)
$$
which establishes (B.49A). Through the interchange of the
subscripts $m$ and $l$, we also obtain (B.50A).

Next, we turn to Case $(B)$  with the boundary condition
$f_n(0)=1$ for all $n$. Therefore at $r=0$,
$$
\frac{f_{m+1}(0)}{f_{l+1}(0)}=1.
\eqno(B.61)
$$
Furthermore from (\ref{e4.15}) and (4.16B), we also have
$f'_{m+1}(0)=f'_{l+1}(0)=0$ and ${\cal D}_{m+1}(0)={\cal D}_{l+1}(0)=0$, with
their ratio given by
\begin{eqnarray}
~~~~~~~~~~~~~~~~\left(\frac{df_{m+1}}{df_{l+1}}\right)_{r=0}
&=&\frac{f'_{m+1}(0)}{f'_{l+1}(0)}
=\frac{{\cal D}_{m+1}(0)}{{\cal D}_{l+1}(0)}
=\frac{{\cal D}'_{m+1}(0)}{{\cal D}'_{l+1}(0)}\nonumber\\
&=& \frac{h(0)-{\cal E}_{m+1}}{h(0)-{\cal E}_{l+1}}.
~~~~~~~~~~~~~~~~~~~~~~~~~~~~~~~~~(B.62)\nonumber
\end{eqnarray}
From (B.1) of Lemma 1, we see that if $(f_m/f_l)'<0$,
then ${\cal E}_{m+1}>{\cal E}_{l+1}$ and therefore
$$
\left(\frac{df_{m+1}}{df_{l+1}}\right)_{r=0}<1,
\eqno(B.63)
$$
$$
\left( \frac{\hat{\eta}}{\hat{\xi}}\right)_{r=0}=1.
\eqno(B.64)
$$
Thus,
$$
\left( \frac{d\hat{\eta}}{d\hat{\xi}}\right)_{r=0}<\left(
\frac{\hat{\eta}}{\hat{\xi}}\right)_{r=0}.
\eqno(B.65)
$$
Analogously to (B.18), define
$$
L(r) \equiv \hat{\xi} \frac{d\hat{\eta}}{d\hat{\xi}} - \hat{\eta}
=f_{l+1}(r) \frac{f'_{m+1}(r)}{f'_{l+1}(r)}-f_{m+1}(r);
\eqno(B.66)
$$
therefore
\begin{eqnarray*}
~~~~~~~\frac{dL(r)}{dr}&=&\hat{\xi}' \frac{dL}{d\hat{\xi}}
=\hat{\xi}'\hat{\xi}
\frac{d}{d\hat{\xi}}\frac{d\hat{\eta}}{d\hat{\xi}} \\
&=&\hat{\xi}
\frac{d}{dr}\left(\frac{d\hat{\eta}}{d\hat{\xi}}\right)=
f_{l+1}\frac{d}{dr}\left(\frac{f'_{m+1}}{f'_{l+1}} \right)
=f_{l+1}\frac{d}{dr}\left(\frac{{\cal D}_{m+1}}{{\cal D}_{l+1}} \right).
~~~~~(B.67)
\end{eqnarray*}
From (B.21) of Lemma 3, we know    that if $(f_m/f_l)'<0$
then $({\cal D}_{m+1}/{\cal D}_{l+1})'<0$, which leads to
$$
\frac{dL(r)}{dr}<0.
\eqno(B.68)
$$
From (B.66), we have
$$
L(r) = \hat{\xi}\left( \frac{d\hat{\eta}}{d\hat{\xi}} - \frac{
\hat{\eta}}{\hat{\xi}} \right) =f_{l+1}(r) \left(
\frac{d\hat{\eta}}{d\hat{\xi}} - \frac{ \hat{\eta}}{\hat{\xi}}
\right),
\eqno(B.69)
$$
and therefore at $r=0$, because of (B.65),
$$
L(0)<0.
\eqno(B.70)
$$
Combining (B.68) and (B.70), we derive
$$
L(r)<0~~~~~~~{\sf for}~~~r\geq 0.
\eqno(B.71)
$$

Multiplying (B.66) by $f'_{l+1}(r)$, we have
\begin{eqnarray*}
~~~~~~~~~~~~~~~f'_{l+1}(r) L(r) &=&f_{l+1}(r)
f'_{m+1}(r)-f_{m+1}(r)f'_{l+1}(r)\nonumber\\
&=& f^2_{l+1}(r)\left(\frac{f_{m+1}(r)}{f_{l+1}(r)}\right)'.
~~~~~~~~~~~~~~~~~~~~~~~~~~(B.72)
\end{eqnarray*}
Because $f'_{l+1}(r)$ and $L(r)$ are  both negative, it follows
then
$$
\left(\frac{f_{m+1}(r)}{f_{l+1}(r)}\right)'>0,
$$
which gives (B.49B) for Case $(B)$, with  the boundary condition
$f_n(0)=1$. Interchanging the subscripts $m$ and $l$, (B.49B)
becomes (B.50B), and Lemma 4 is established.

We now turn to the proof of the theorem stated in
(\ref{e4.26})-(\ref{e4.34}).\\

\noindent \underline{Proof of the Hierarchy Theorem}

When $n=0$, we have
$$
f_0(r)=1.
\eqno(B.73)
$$
From (\ref{e4.22})-(\ref{e4.24}), we find for $n=1$
$$
f'_1(r)<0,
\eqno(B.74)
$$
and therefore
$$
(f_1/f_0)'<0.
\eqno(B.75)
$$

In Case $(A)$, by using (B.49A) and by setting $m=1$ and $l=0$, we
derive $(f_2/f_1)'<0$; through induction, it follows then
$(f_{n+1}/f_n)'<0$ for all $n$. From Lemma 1, we also find ${\cal
E}_{n+1}>{\cal E}_n$ for all n. Thus,
(\ref{e4.26})-(\ref{e4.29}) are established.

In Case $(B)$, by using (B.75) and (B.49B), and setting
$m=1$ and $l=0$, we find $(f_2/f_1)'>0$, which in turn leads to
$(f_3/f_2)'<0$, $\cdots$, and (\ref{e4.33})-(\ref{e4.34}).
Inequalities (\ref{e4.30})-(\ref{e4.32}) now follow from
(B.1)-(B.2) of Lemma 1. The Hierarchy Theorem is
proved.

Assuming that $h(0)$ is finite, we have for any $n$
$$
0<{\cal E}_n<h(0). \eqno(B.76)
$$
Therefore, each of the monotonic sequences
$$
{\cal E}_1<{\cal E}_2<{\cal E}_3<\cdots,~~~~{\sf in}~~~(A)
$$
$$
{\cal E}_1<{\cal E}_3<{\cal E}_5<\cdots,~~~~{\sf in}~~~(B)
$$
and
$$
{\cal E}_2>{\cal E}_4>{\cal E}_6>\cdots,~~~~{\sf in}~~~(B)
$$
converges to a finite limit ${\cal E}$. By following the
discussions in Section~5 of Ref.[18], one can show that each of the
corresponding monotonic sequences of $f_n(r)$ also converges to a
finite limit $f(r)$. The interchange of the limit $n\rightarrow
\infty$ and the integrations in (4.16A) completes the proof that
in Case (A) the limits ${\cal E}$ and $f(r)$ satisfy
$$
f(r)=1 - 2 \int_{r}^{\infty}\chi^{-2}(y)dy
\int_{y}^{\infty}\chi^{2}(z) (h(z)- {\cal E}) f(z) dz.
\eqno(B.77A)
$$
As noted before, the convergence in Case (A) can hold for any
large but finite and positive $h(r)$, provided that $h'(r)$ is negative for
$r>0$. In Case (B), a large $h(r)$ may yield a negative $f_n(r)$,
in violation of (4.25B) . Therefore, the convergence does depend
on the smallness of $h(r)$. One has to follow discussions similar
to those given in Ref.[17] to ensure that the limits ${\cal E}$ and
$f(r)$ satisfy
$$
f(r)=1 - 2 \int_{0}^{r}\chi^{-2}(y)dy \int_0^{y}\chi^{2}(z) (h(z)-
{\cal E}) f(z) dz. \eqno(B.77B)
$$

\newpage

\section*{\bf Appendix C.~~~ Comparison with Perturbative\\
\hspace*{4.5cm} Expansion } \setcounter{section}{9}
\setcounter{equation}{0}

\noindent{\bf C.1 A simple Model and Its Iterative Solutions}\\

Consider the one-dimensional problem in which $V$, $U$ and $h$ of
(\ref{e2.1}), (\ref{e2.7}) and (\ref{e2.11}) are given by
\begin{eqnarray*}
~~~V(x)=V(-x)= \left\{\begin{array}{lcl}
\infty&&~~~~~|x|>L+l\\
0&~~~~~~~~{\sf for}~~&l<|x|<L+l,\\
-\frac{1}{2}g^2&&~~~~~|x|<l,
\end{array}
\right.~~~~~~~~~~~~~~~~(C.1)
\end{eqnarray*}
\begin{eqnarray*}
~~~U(x)=U(-x)= \left\{\begin{array}{lcl}
\infty&&~~~~~~~~~~|x|>L+l,\\
 &~~~~~~~~~~~~{\sf for} \\
0&&~~~~~~~~~~|x|<L+l
\end{array}
\right.~~~~~~~~~~~~~(C.2)
\end{eqnarray*}
and
\begin{eqnarray*}
~~~h(x)=U(x)-V(x)= \left\{\begin{array}{lcl}
0&&~~~~~~~~~~|x|>l,\\
 &~~~~~~~~~{\sf for} \\
\frac{1}{2}g^2&&~~~~~~~~~~|x|<l.
\end{array}
\right.~~~~~~~~~~~~~(C.3)
\end{eqnarray*}
Let $\phi(x)$ and $\psi(x)$ be the respective groundstates of
$$
H_0\phi=-\frac{1}{2}\frac{d^2\phi}{dx^2}+U\phi=E_0\phi
\eqno(C.4)
$$
and
$$
H\psi=-\frac{1}{2}\frac{d^2\psi}{dx^2}+V\psi=E\psi. \eqno(C.5)
$$
As a model, (C.5) is the Schroedinger equation that we wish to
solve, and (C.4) is the equation that the trial function $\phi$
satisfies. The iterative equation (\ref{e2.19}) becomes
$$
(-\frac{1}{2}\frac{d^2}{dx^2}+U(x)-E_0)\psi_n(x)= (h(x)-{\cal
E}_n)\psi_{n-1}(x). \eqno(C.6)
$$
with $\psi_0=\phi(x)$. Since $U$ and $V$ are both even in $x$, we
need only to consider the region
$$
x\geq 0. \eqno(C.7)
$$

Set within $x<L+l$
$$
\phi(x)=\cos px, \eqno(C.8)
$$
with
$$
p(L+l)=\frac{\pi}{2}~~~~{\sf and}~~~~E_0=\frac{p^2}{2}. \eqno(C.9)
$$
Introduce, as in (\ref{e2.25}), $f_n=\psi_n/\phi$. We have in
accordance with (\ref{e3.1})-(\ref{e3.7})
$$
D_n(x)=-\frac{1}{2}\kappa(x)\frac{df_n}{dx},~~~\kappa(x)=\phi^2(x),
\eqno(C.10)
$$
$$
\frac{dD_n}{dx}=\sigma_n(x)=(h(x)-{\cal E}_n)\phi^2(x)f_{n-1}(x),
\eqno(C.11)
$$
$$
\int\limits_0^\infty \sigma_n(x)dx=0 \eqno(C.12)
$$
and therefore
$$
{\cal E}_n=\frac{\int\limits_0^\infty
h\phi^2f_{n-1}dx}{\int\limits_0^\infty\phi^2f_{n-1}dx}~.
\eqno(C.13)
$$

One can verify that for $n=1$
$$
{\cal E}_1=\frac{g^2}{2(L+l)}(l+\frac{1}{2p}\sin 2pl), \eqno(C.14)
$$
\begin{eqnarray*}
D_1(x)=\frac{g^2}{4(L+l)} \left\{\begin{array}{ll}
(l+\frac{\sin 2pl}{2p})\bigg(L+l-x-\frac{\sin 2px}{2p}\bigg),&~l<x<L+l,\\
(x+\frac{\sin 2px}{2p})\bigg(L-\frac{\sin 2pl}{2p}\bigg),&~0<x<l
\end{array}
\right.(C.15)
\end{eqnarray*}
and
\begin{eqnarray*}
f_1(x)= \left\{\begin{array}{lcl}
f_1(R)+\frac{{\cal E}_1}{p^2}[1-p(L+l-x)\tan px],&&~~~l<x<L+l,\\
f_1(0)-\frac{1}{p}(\frac{g^2}{2}-{\cal E}_1)x \tan px,&&~~~0<x<l.
\end{array}
\right.~~(C.16)
\end{eqnarray*}
The constants $f_1(R)$ and $f_1(0)$ are related through the
continuity of $f_1(x)$ at $x=l$. In order to derive the general
solution for $f_n(x)$, it is useful to consider a limited region
in $x$ (either $|x|<l$ or $|x|>l$) over which
$$
\Delta_n \equiv 2(h(x)-{\cal E}_n)={\sf constant}. \eqno(C.17)
$$
From (C.10)-(C.11), we have
$$
-\frac{d}{dx}\bigg(\cos ^2 px\frac{df_n}{dx}\bigg) =\Delta_n \cos
^2 px f_{n-1}. \eqno(C.18)
$$
Write
$$
f_n=P_n(x)+ (\tan px) Q_n(x). \eqno(C.19)
$$
We find, from (C.18),
$$
\frac{d^2}{dx^2}P_n+2p\frac{d}{dx}Q_n=-\Delta_n P_{n-1}
$$
$$
{\sf and}~~~~~~~~~~~~~~~~~~~~~~~~~~~~~~~~~~~~~~~~~~~~~~~~~~
~~~~~~~~~~~~~~~~~~~~~~~~~~~~~~~~~~~~~~~~~~~~~ \eqno(C.20)
$$
$$
\frac{d^2}{dx^2}Q_n-2p\frac{d}{dx}P_n=-\Delta_n Q_{n-1}.
$$
Thus, $P_n$ and $Q_n$ are both $n^{th}$ order polynomials in $x$.
Define
$$
P^\pm_n \equiv \frac{1}{2}(P_n\mp iQ_n)=(P_n^\mp)^*~. \eqno(C.21)
$$
From (C.20), $P_n^\pm$ satisfies
$$
(\frac{d^2}{dx^2}\pm 2ip\frac{d}{dx})P_n^\pm =-\Delta_n
P_{n-1}^\pm. \eqno(C.22)
$$
It is useful to introduce a set of $n^{th}$ order polynomials
$R_n^\pm$, independent of $\Delta_n$. We require
$$
(\frac{d^2}{dx^2}\pm 2ip\frac{d}{dx})R_n^\pm =-R_{n-1}^\pm
\eqno(C.23)
$$
with the boundary conditions (i) for $n=0$,
$$
R_0^\pm(x)=1~~~{\sf at~all}~~x, \eqno(C.24)
$$
and (ii) for $n>0$, at $x=0$
$$
R_n^\pm(0)=0. \eqno(C.25)
$$
The solution for $R_n^\pm(x)$ can be further simplified by
introducing a set of $n^{th}$ order polynomials $R_n(x|q)$ without
the superscript $+$ or $-$, but depending on a scaling parameter
$q$, so that
$$
R_n(x|q)=q^{2n} R_n(\frac{x}{q}|1) \eqno(C.26)
$$
where $R_n(\frac{x}{q}|1)$ is an $n^{th}$ order polynomial of
$\frac{x}{q}$ with constant coefficients. The $R_n(x|q)$ are
related to $R_n^\pm(x)$ defined by (C.23)-(C.25) through
\begin{eqnarray*}
\left.\begin{array}{lcl}
~~~~~~~~~~~~~R_n(x|q)&=&R_n^+(x)~~~{\sf when}~~~q=q^+ \equiv +\frac{i}{2p}\\
~~~~~~~~~~~~~R_n(x|q)&=&R_n^-(x)~~~{\sf when}~~~q=q^- \equiv
-\frac{i}{2p}.
\end{array}
\right.~~~~~~~~~~~~~~~~(C.27)
\end{eqnarray*}
Writing
$$
u \equiv \frac{x}{q}~~~{\sf and}~~~R_n(\frac{x}{q}|1) \equiv
r_n(u), \eqno(C.28)
$$
we have
$$
r_0(u)=1,~~~r_1(u)=u,~~~r_2(u)=\frac{1}{2}u^2+u,
$$
$$
r_3(u)=\frac{1}{6}u^3+u^2+2u, \eqno(C.29)
$$
$$
r_4(u)=\frac{1}{24}u^4 +\frac{1}{2}u^3+\frac{5}{2}u^2+5u,~~~{\sf
etc.;}
$$
correspondingly
$$
R_0(x|q)=1,~~~~R_1(x|q)=qx,~~~~R_2(x|q)=\frac{1}{2}q^2x^2+q^3x,
$$
$$
R_3(x|q)=\frac{1}{6}q^3x^3+q^4x^2+2q^5x, \eqno(C.30)
$$
$$
R_4(x|q)=\frac{1}{24}q^4x^4+\frac{1}{2}q^5x^3+\frac{5}{2}q^6x^2+5q^7x,~{\sf
etc.}
$$

In terms of $R_n^\pm(x)$, the $P_n^\pm(x)$ that satisfies (C.22)
is given by
\begin{eqnarray*}
~~~~~~~~~P_n^\pm(x)&=&\frac{1}{2}\Delta_1\Delta_2 \cdots\Delta_n R_n^\pm(x)\\
&&+c_1^\pm \Delta_2\Delta_3 \cdots\Delta_n R_{n-1}^\pm(x)\\
&&+c_2^\pm \Delta_3 \cdots\Delta_n R_{n-2}^\pm(x)+\cdots\\
&&+c_{n-1}^\pm \Delta_n R_1^\pm(x) +c_n^\pm ~~~~~~~~~~~~~~~~~~~~~~
~~~~~~~~~~~~~~~(C.31)\\
\end{eqnarray*}
where $c_1^\pm,~c_2^\pm,~\cdots,c_n^\pm$ are integration
constants, with $c_m^\pm=(c_m^\mp)^*$ for all $m$. Thus for $n=1$,
since $R_1(x|q)=qx$,
$$
R_1^\pm(x) = R_1(x|q^\pm)=R_1(x|\pm\frac{i}{2p})=\pm\frac{ix}{2p}.
\eqno(C.32)
$$
According to (C.31),
$$
P_1^\pm(x)=\frac{1}{2} \Delta_1 R_1^\pm(x) +c_1^\pm=
\pm\frac{ix}{4p}\Delta_1+c_1^\pm \eqno(C.33)
$$
and from (C.21)
$$
P_1=P_1^++P_1^-=c_1^++c_1^-~~~~~~~~~~~~~~~~~~~~~
$$
$$
{\sf and}~~~~~~~~~~~~~~~~~~~~~~~~~~~~~~~~~~~~~~~
~~~~~~~~~~~~~~~~~~~~~~~~~~~~~~~~~~~~~~~~~~~~~~~~ \eqno(C.34)
$$
$$
Q_1=i(P_1^+-P_1^-)=-\frac{x}{2p}\Delta_1+i(c_1^+-c_1^-).
$$
Since
\begin{eqnarray*}
~~~\Delta_1=2(h(x)-{\cal E}_1)= \left\{\begin{array}{lcl}
-2{\cal E}_1&&~~~~~l<x<L+l,\\
 &~~~~~~~~~{\sf for} \\
g^2-2{\cal E}_1&&~~~~0<x<l,
\end{array}
\right.~~(C.35)
\end{eqnarray*}
we have, in agreement with (C.16), $P_1$ is a constant in
$l<x<L+l$, and a different constant in $0<x<l$. Likewise,
$$
Q_1=\frac{{\cal E}_1x}{p} + {\sf
constant~~~~~~~~~~~~~~~~~~~~~~~~~~~~~~~~in}~l<x<L+l
$$
$$
{\sf and} ~~~~~~~~~~~~~~~~~~~~~~~~~~~~~~~~~~~~~~~~~~~~~~~~~~~~~
~~~~~~~~~~~~~~~~~~~~~~~~~~~~~~~~~~~~~~~~~~~~~~~~~~~~~\eqno(C.36)
$$
$$
Q_1=-\frac{1}{p}(\frac{g^2}{2}-{\cal E}_1)x +~{\sf a~different~
constant~~~~~~~~in}~ 0<x<l.
$$
These integration constants are determined by the continuity
equations between regions, leaving one final overall constant that
is not determined. Likewise, we can derive the solution $f_n(x)$
for any $n$.

If we impose the boundary condition at $x=L+l$,
$$
f_n(L+l)=1,  \eqno(C.37)
$$
then in accordance with the Hierarchy Theorem
$$
{\cal E}_1 < {\cal E}_2 < \cdots  \eqno(C.38)
$$
with
$$
\lim\limits_{n\rightarrow \infty} {\cal E}_n={\cal E}=E_0-E
\eqno(C.39)
$$
for any finite $g^2$.

\newpage

\noindent{\bf C.2 Analyticity of $E(g^2)$}\\

The convergence of the perturbative series in $g^2$ depends on, in
the complex $g^2$ plane, the location of the nearest singularity
$g_c^2$ of $E(g^2)$ to the origin. While the simple model
discussed above can be readily solved, it is more complicated to
locate the singularities of $E(g^2)$. We first discuss the exact
solution of (C.5). Again we need only consider the positive
$x$-axis. Note that the well-depth $-\frac{g^2}{2}$ has a critical
value $-\frac{g_0^2}{2}$; for $g^2>g^2_0$, $E<0$, and for
$g^2<g_0^2$, $E>0$. When $g^2=g_0^2$, $E=0$ and
\begin{eqnarray*}
~~~~~~~~~~~~~\psi=\psi_c \propto \left\{\begin{array}{lcl}
\frac{L+l-x}{L}& &~~~~~l<x<L+l,~~~\\
&~~~~~~~~~~~~~{\sf for}&\\
 \cos q_0x& &~~~~0<x<l~~~
\end{array}
\right.~~(C.40)
\end{eqnarray*}
with
$$
q_0 \tan q_0l=\frac{1}{L}, \eqno(C.41)
$$
as required by the continuity of $\psi'/\psi$ at $x=l$. When
$g^2<g_0^2$,
\begin{eqnarray*}
~~~~~~~~~~~~\psi \propto \left\{\begin{array}{lcl}
\sin [k(L+l-x)]& &~~~~l<x<L+l,\\
&~~~~~~~~~~~{\sf for}&\\
\cos qx& &~~~~0<x<l~~~~~~~
\end{array}
\right.~~(C.42)
\end{eqnarray*}
with
$$
q \tan ql=k \cot kL, \eqno(C.43)
$$
and
$$
E=\frac{k^2}{2}=\frac{1}{2}(q^2-g^2)>0. \eqno(C.44)
$$
When $g^2>g_0^2$,
\begin{eqnarray*}
~~~~~~~~~~~\psi \propto \left\{\begin{array}{lcl}
\sinh [\kappa(L+l-x)]& &~~~~l<x<L+l,\\
&~~~~~~~~~~~{\sf for}&\\
\cos qx& &~~~~0<x<l~~~~
\end{array}
\right.~~(C.45)
\end{eqnarray*}
with
$$
q \tan ql=\kappa \coth \kappa L \eqno(C.46)
$$
and
$$
E=-\frac{\kappa^2}{2}=\frac{1}{2}(q^2-g^2)<0. \eqno(C.47)
$$
In the limit $L\rightarrow\infty $, $q_0=0$, in accordance with
(C.41), and therefore
$$
g_0^2=0. \eqno(C.48)
$$
For any $g^2$ however small, $E$ is negative. Solution (C.45)
becomes
\begin{eqnarray*}
~~~~~~~~~~~~~~~~~~~~~\psi \propto \left\{\begin{array}{lcl}
e^{-\kappa x}& &~~~~x>l,~~~~~~~~~~\\
&~~~~~~~~~~~~~~~~~{\sf for}&\\
\cos qx& &~~~~x<l
\end{array}
\right.~~(C.49)
\end{eqnarray*}
with
$$
E=-\frac{\kappa^2}{2}=-\frac{1}{2}(g^2-q^2) \eqno(C.50)
$$
and
$$
\kappa = q \tan ql. \eqno(C.51)
$$

We examine the analyticity of $E(g^2)$ for the case $L=\infty$.
the same analysis can be extended to arbitrary $L$, but it is more
complex. The singularities of $E=E(g^2)$ can be determined by
setting $\frac{dE}{dg^2}=\infty$; i.e., $\frac{dg^2}{dE}=0$, which
leads to, on account of $E=-\frac{1}{2}\kappa^2$,
$$
\frac{dg^2}{d\kappa^2}=0. \eqno(C.52)
$$
Because $g^2=q^2+\kappa^2$ according to (C.50), (C.52) is
equivalent to $\frac{dq^2}{d\kappa^2}+1=0$; i.e.,
$$
\frac{dq}{d\kappa}=-\frac{\kappa}{q}. \eqno(C.53)
$$
Differentiating (C.51), we find
$$
\frac{d\kappa}{dq} = \tan ql+ql\sec^2 ql, \eqno(C.54)
$$
which, when combined with (C.53), gives
$$
-\frac{q}{\kappa} = \tan ql+ql\sec^2 ql. \eqno(C.55)
$$
Since $(q/\kappa)=\cot ql$ according to (C.51), (C.55) becomes
$$
\cot ql+ \tan ql+ql\sec^2 ql=0;
$$
i.e.,
$$
\sec^2 ql(\cot ql+ ql)=0.
$$
When $\sec^2 ql=0$, $ql=\pm i\infty$ which is uninteresting for
our purpose. Hence, for singularities at a finite $g^2$, we
require
$$
\cot ql+ ql=0, \eqno(C.56)
$$
i.e.,
$$
ql \tan ql=-1. \eqno(C.57)
$$
Define
$$
z= ql. \eqno(C.58)
$$
Our problem is then reduced to the study of zeroes of
$$
z \tan z+1=0. \eqno(C.59)
$$
For each zero at $z$, there is a singularity of $E(g^2)$ in the
complex $g^2$ plane at $g^2=\kappa^2+q^2=q^2 \tan^2 ql+q^2$; i.e.,
at
$$
g^2l^2=z^2 \tan^2 z+z^2=1+z^2. \eqno(C.60)
$$
(i) For $z$ imaginary, we set $z=\pm iy$. (C.59) becomes $y \tanh
y=1$, which gives for the smallest $|y|\cong 1.2$, $z\cong \pm 1.2
i$ and
$$
g^2l^2\cong -0.44~. \eqno(C.61)
$$
As will be shown below, this gives $g_c^2$.

\noindent(ii) For $z$ real, the nearest zeroes for (C.59) are
$z\cong \pm 2.8$ and correspondingly
$$
g^2l^2\cong 8.8~. \eqno(C.62)
$$
(iii) We shall now establish that in the complex $g^2$-plane, the
nearest singularity $g_c^2$ of $E(g^2)$ to the origin is given by
(C.61); i.e.,
$$
g_c^2l^2\cong -0.44~. \eqno(C.63)
$$
Consider the well known expansion
$$
\tan
z=2z\sum\limits_{n=0}^\infty\frac{1}{(n+\frac{1}{2})^2\pi^2-z^2}~.
$$
For $z$ near $\frac{\pi}{2}$, we may write the righthand side as
\begin{eqnarray*}
&&2z\bigg[\frac{1}{(\frac{\pi}{2})^2-z^2}+
\sum\limits_{n=1}^\infty\frac{1}{(n+\frac{1}{2})^2\pi^2-z^2}\bigg]\\
&=&2z\bigg[\frac{1}{(\frac{\pi}{2})^2-z^2}+
\sum\limits_{n=1}^\infty\bigg(\frac{1}{(n+\frac{1}{2})^2\pi^2}+
\frac{z^2}{(n+\frac{1}{2})^4\pi^4}+\cdots\bigg)\bigg]\\
&=&2z\bigg[\frac{1}{(\frac{\pi}{2})^2-z^2}+(\frac{1}{2}-\frac{4}{\pi^2})+
z^2\sum\limits_{n=1}^\infty\frac{1}{(n+\frac{1}{2})^4\pi^4}+\cdots\bigg].
~~~~~~~~~~~~~~(C.64)
\end{eqnarray*}
By using first the approximation
$$
\tan z\cong
2z\bigg[\frac{1}{(\frac{\pi}{2})^2-z^2}+(\frac{1}{2}-\frac{4}{\pi^2})\bigg]~
\eqno(C.65)
$$
and then the correction due to the successive remaining terms in
(C.64), one can readily show that (C.61) gives the smallest
singularity $g_c^2$ in the $g^2$-plane. (Note that in this case,
because $L=\infty$, $g_0=0$ in accordance with (C.48).) Thus, the
perturbative series of $E(g^2)$ is convergent only if
$g^2l^2<|g_c^2|l^2\cong 0.44~$. On the other hand, the iterative
series of Section C.1 is convergent for any finite $g^2$.\\

\noindent{\bf C.3 A Comparative Study of Different Green's Functions}\\

\noindent(i) Write the iterative equation (\ref{e2.19}) as
$$
(H_0-E_0)\psi_n(q)=S_n(q) \eqno(C.66)
$$
where
$$
S_n(q)=(h(q)-{\cal E}_n)\psi_{n-1}(q). \eqno(C.67)
$$
In one dimension, setting $q=x$ and using (\ref{e2.25}) and
(\ref{e3.15})-(\ref{e3.16}), the solution $\psi_n(x)$ is given by
$$
\psi_n(x)=\psi_n(\infty)+\int\limits_{-\infty}^\infty (x|{\cal
G}|z)S_n(z)dz \eqno(C.68)
$$
where the Green's function ${\cal G}$ is
$$
(x|{\cal G}|z)=-2\phi(x)\int\limits_{x}^\infty \phi^{-2}(y)dy
\theta (y-z)\phi(z)   \eqno(C.69)
$$
with
\begin{eqnarray*}
~~~~~~~~~~~~~~~~\theta (y-z)= \left\{\begin{array}{lcl}
1& &~~~~z>y,~~~~~~~~~~\\
&~~~~~~~~~~~~~~~~~~{\sf for}&\\
0& &~~~~z<y,
\end{array}
\right.~~~(C.70)
\end{eqnarray*}
and therefore
$$
(x|{\cal G}|z)=0~~~~~{\sf if}~~~~z<x.   \eqno(C.71)
$$
The function $(x|{\cal G}|z)$ is one of the Sturm-Liouville family
of Green's functions[20]. On the other hand, in deriving the usual
perturbative series one deals often with similar equations of the
form
$$
(H_0-\lambda)u(q)=v(q) \eqno(C.72)
$$
where $\lambda$ is a parameter. Let $u_n$ be the $n^{th}$ member
of the ortho-normal set of real eigenfunctions of $H_0$ with
eigenvalue $e_n$. We assume
\begin{eqnarray*}
~~~~~~~~~~~\int u_n(q)u_m(q)d^Nq = \delta_{nm}=
\left\{\begin{array}{lcl}
0,& &~~~~n\neq m,~~~~~\\
1,& &~~~~n=m.
\end{array}
\right.~~~~~~~~~~~~(C.73)
\end{eqnarray*}
Hence,
$$
(H_0-e_n)u_n(q)=0 \eqno(C.74)
$$
which reduces to (\ref{e2.8}) when $n=0$; i.e.,
$$
e_0=E_0~~~~{\sf and}~~~~u_0=\phi. \eqno(C.75)
$$
For $\lambda$ different from any of the eigenvalues $e_n$, the
general solution of (C.72) is
$$
u(q)= \int
(q|G(\lambda)|\overline{q})v(\overline{q})d^N\overline{q}
\eqno(C.76)
$$
where
$$
(q|G(\lambda)|\overline{q})=\sum\limits_n\frac{u_n(q)u_n(\overline{q})}{e_n-\lambda}~.
\eqno(C.77)
$$
To simplify our notations, we assume all $u_n(q)$ to be real. When
$\lambda=E_0=e_0$, (C.72) becomes
$$
(H_0-e_0)u(q)=v(q), \eqno(C.78)
$$
same as (C.66). Since $(H_0-e_0)u_0=0$, we have
$$
\int u_0(q)v(q)d^Nq=0.  \eqno(C.79)
$$
The general solution of (C.78) is
$$
u(q)= \int
(q|\hat{G}(\lambda)|\overline{q})v(\overline{q})d^N\overline{q}+cu_0(q)
\eqno(C.80)
$$
with $c$ an arbitrary constant and
$$
(q|\hat{G}(e_0)|\overline{q}) \equiv \sum\limits_{n\neq 0}
\frac{u_n(q)u_n(\overline{q})}{e_n-e_0}~. \eqno(C.81)
$$
In the sub-section (iii) below, we shall analyze the difference
between ${\cal G}$ and $G$, or $\hat{G}$. Before that, we shall
review the roles of $G$ and $\hat{G}$ in the usual perturbative
formulas..

\noindent(ii) In order to solve the Schroedinger equation
(\ref{e2.4}) perturbatively, we begin with (\ref{e2.13}); i.e.,
$$
(H_0-E_0)\psi(q)=(h-{\cal E})\psi(q). \eqno(C.82)
$$
Following the standard Brillouin-Wigner procedure, we require
$$
\int u_0(q)\psi(q)d^Nq= \int \phi(q)\psi(q)d^Nq=1. \eqno(C.83)
$$
Define
$$
\hat{\psi}(q)=\psi(q)-u_0(q), \eqno(C.84)
$$
which is orthogonal to $u_0(q)$, because of (C.83). Eq.(C.82)
leads to
$$
(H_0-E_0)\hat{\psi}(q)=(h-{\cal E})(u_0(q)+\hat{\psi}(q));
\eqno(C.85)
$$
on account of (\ref{e2.12}), this gives
$$
(H_0-E)\hat{\psi}(q)=(h-{\cal E})u_0(q)+h\hat{\psi}(q).
\eqno(C.86)
$$
Multiply (C.86) on the left by $u_n$ and integrate. We have for
$n=0$
$$
 {\cal E}= \int u_0  h (u_0+\hat{\psi})d^Nq,
\eqno(C.87)
$$
the same as (\ref{e2.16}); for $n\neq0$
$$
(e_n-E)(n|\hat{\psi})= \int u_n  h (u_0+\hat{\psi})d^Nq
\eqno(C.88)
$$
where
$$
(n|\hat{\psi})\equiv \int u_n \hat{\psi}d^Nq. \eqno(C.89)
$$
Introduce the sequence
$$
 \hat{\psi}_1,~ \hat{\psi}_2,~ \hat{\psi}_3,~\cdots \eqno(C.90)
$$
with
$$
(0|\hat{\psi}_l)=0~~~{\sf for~~all}~~~l\geq 1. \eqno(C.91)
$$
When  $l=1$, we require for all $n\neq 0$
$$
(e_n-E)(n|\hat{\psi}_1)= \int u_n  h u_0 d^Nq.  \eqno(C92)
$$
For $l>1$,
$$
(e_n-E)(n|\hat{\psi}_l)= \int u_n  h \hat{\psi}_{l-1} d^Nq.
\eqno(C.93)
$$
Therefore, the sum
$$
\hat{\psi}= \sum\limits_1^\infty \hat{\psi}_l  \eqno(C.94)
$$
satisfies (C.86). In compact form,
$$
\hat{\psi}(q)= \int (q|\hat{G}(E)|\overline{q})
\hat{S}(\overline{q}) d^N\overline{q} \eqno(C.95)
$$
where
$$
\hat{S}(q)=h \hat{\psi}(q) +(h-{\cal E})u_0(q). \eqno(C.96)
$$
The usual perturbative series can then be readily derived by first
introducing a parameter $\epsilon$, defined by
$$
h(q)=\epsilon \hat{h}(q). \eqno(C.97)
$$
The subsequent expansions of
$$
E=E_0+\epsilon E(1)+\epsilon^2 E(2)+\cdots \eqno(C.98)
$$
$$
\hat{\psi}_l= \epsilon \hat{\psi}_l(1)+\epsilon^2
\hat{\psi}_l(2)+\cdots \eqno(C.99)
$$
and
\begin{eqnarray*}
~~~~~~~\hat{G}(E)&=&\hat{G}(E_0)+\epsilon E(1)\frac{\partial
\hat{G}(E_0)}{\partial E_0}\\
&&+\epsilon^2 \bigg[E(2)\frac{\partial }{\partial
E_0}+\frac{1}{2}(E(1))^2\frac{\partial^2 }{\partial
E_0^2}\bigg]\hat{G}(E_0)+\cdots ~~~~~~~~~~(C.100)
\end{eqnarray*}
lead to the usual perturbative series.\\


\noindent(iii) According to (C.68)-(C.69), the Green's function
${\cal G}$ satisfies
$$
(H_0-E_0) (x|{\cal G}|z)= \delta(x-z), \eqno(C.101)
$$
whereas the function $\hat{G}(E_0)$ given by (C.81) satisfies, in
a one-dimensional problem,
$$
(H_0-E_0) (x|\hat{G}(E_0)|z)= \delta(x-z)-u_0(x)u_0(z)
\eqno(C.102)
$$
with $u_0=\phi$ in accordance with (C.75). Thus ${\cal G}$ and
$\hat{G}$ are clearly different. As an explicit example, let
$$
H_0=-\frac{1}{2}\frac{d^2}{dx^2}+U(x) \eqno(C.103)
$$
and
\begin{eqnarray*}
~~~~~~~~~~~~~~U(x) = \left\{\begin{array}{lcl}
\infty&&~~~~~x>R\\
0&~~~~~~~~~~~~~~~~~~~~{\sf for}~~&0<x<R,\\
\infty&&~~~~~x<0,
\end{array}
\right.~~~~~~~~~~~(C.104)
\end{eqnarray*}
similar to (C.2) and (C.4). The corresponding eigenfunction
$u_n(x)$ and eigenvalue $e_n$ that satisfy (C.73)-(C.74) are
$$
u_n=\sqrt{\frac{2}{R}} \sin k_nx~~~{\sf
and}~~~e_n=\frac{1}{2}k_n^2,~~~{\sf with}~~~k_n=\frac{n+1}{R}\pi,
\eqno(C.105)
$$
and $n=0,~1,~2,~\cdots$. For $\lambda$ different from any of the
eigenvalues $e_n$, $G(\lambda)$ of (C.77) in this one-dimensional
problem is given by
$$
(x|G(\lambda)|z)=\sum\limits_{n=0}^\infty \frac{1}{e_n-\lambda}
u_n(x)u_n(z) \eqno(C.106)
$$
where $x$ and $z$ can vary between $0$ and $R$. Introduce
$$
\chi_+= \sin p(R-x),~~~\chi_-= \sin px, \eqno(C.107)
$$
$$
\lambda=\frac{1}{2}p^2 \eqno(C.108)
$$
and
$$
\omega=\bigg(\frac{d\chi_-}{dx}\bigg)\chi_+ -
\bigg(\frac{d\chi_+}{dx}\bigg)\chi_- =p\sin pR. \eqno(C.109)
$$
The Green's function $G(\lambda)$ can also be written as
\begin{eqnarray*}
~~~~~~~~(x|G(\lambda)|y) = \frac{2}{\omega}
\left\{\begin{array}{lcl}
\chi_-(x)\chi_+(y)& &~~~~x<y,~~~~~\\
&~~~~~{\sf for}&\\
\chi_+(x)\chi_-(y)& &~~~~x>y.
\end{array}
\right.~~~~~~~~(C.110)
\end{eqnarray*}
One can readily verify that $G(\lambda)$ satisfies
$$
(x|G(\lambda)|y)=(y|G(\lambda)|x), \eqno(C.111)
$$
$$
(-\frac{1}{2}\frac{d^2}{dx^2}+U(x)) (x|G(\lambda)|y) = \delta(x-y)
\eqno(C.112)
$$
and the boundary conditions
$$
(x|G(\lambda)|0)=(x|G(\lambda)|R)=0. \eqno(C.113)
$$
As
$$
\lambda\rightarrow e_0=\frac{1}{2}(\frac{\pi}{R})^2, \eqno(C.114)
$$
the Green's function $\hat{G}(e_0)$ of (C.81) is given by
\begin{eqnarray*}
~~~~~~~~~\hat{G}(e_0)&=&\lim\limits_{\lambda\rightarrow e_0}
\bigg(G(\lambda)-\frac{1}{e_0-\lambda} u_0(y)u_0(x)\bigg) \\
&&=\sum\limits_{n\neq 0}\frac{1}{e_n-e_0}u_n(y)u_n(x).
~~~~~~~~~~~~~~~~~~~~~~~~~~~~~~~~(C.115)
\end{eqnarray*}
Define
$$
\phi(x)=\sin k_0x~~~{\sf and}~~~\hat{\phi}(x)=\cos k_0x
\eqno(C.116)
$$
with $k_0=\frac{\pi}{R}$. We find
\begin{eqnarray*}
~~~~~(x|\hat{G}|z)&=&\frac{R}{\pi^2}\phi(x)\phi(z)-\frac{2}{\pi}
\bigg(\phi(x)z\hat{\phi}(z)+\phi(z)x\hat{\phi}(x)\bigg)\\
&&+\frac{2R}{\pi} \cdot \left\{\begin{array}{lcl}
\phi(x)\hat{\phi}(z)& &~~~~x<z,~~~~~\\
&~~~~~~~~~~{\sf for}&\\
\phi(z)\hat{\phi}(x)&&~~~~x>z.
\end{array}
\right.~~~~~(C.117)
\end{eqnarray*}
The function ${\cal G}$ can also be expressed in terms of $\phi$
and $\hat{\phi}$. We have
$$
-\frac{R}{\pi}\hat{\phi}(x)=\phi(x)\int\limits_{R/2}^x
\frac{dy}{\phi^2(y)} \eqno(C.118)
$$
and
\begin{eqnarray*}
~~~~~~~~~(x|{\cal G}|z)=\frac{2R}{\pi} \cdot
\left\{\begin{array}{lcl}
\phi(x)\hat{\phi}(z)-\phi(z)\hat{\phi}(x),& &~~~~x<z,\\
 0,& &~~~~x>z.
\end{array}
\right.~~~~~~~~~(C.119)
\end{eqnarray*}
Thus, ${\cal G}$ is related, but quite different from $\hat{G}$
and $G$.

\end{document}